\documentclass[twocolumn]{aastex6}

\newcommand{\HII}{$\rm H~{\scriptstyle II}$}

\newcommand{\kms}{$\rm km\,s^{-1}$}
\newcommand{\bsnr}{$\rm SNR(4750\,\AA)$}
\newcommand{\rsnr}{$\rm SNR(7450\,\AA)$}
\newcommand{\eza}{$\rm EZ\_Ages$}
\newcommand{\avfe}{${\rm \langle Fe \rangle}$}
\newcommand{\mgfe}{${\rm [MgFe]}$}

\usepackage{graphicx}
\usepackage{natbib}
\usepackage{txfonts}
\usepackage{journals}

\shorttitle{LSS-GAC of M31 GC}
\shortauthors{Chen et al.}

\begin{document}
\title{The LAMOST spectroscopic survey of {star clusters in M31}. \\
II. Metallicities, ages and masses}

\author{Bingqiu Chen\altaffilmark{1,7}, 
 Xiaowei Liu\altaffilmark{1,2}, 
 Maosheng Xiang\altaffilmark{3,7},
 Haibo Yuan\altaffilmark{4},
 Yang Huang\altaffilmark{1}, 
 Jianrong Shi\altaffilmark{3},
 Zhou Fan\altaffilmark{3},
Zhiying Huo\altaffilmark{3},
Chun Wang\altaffilmark{1},
Juanjuan Ren\altaffilmark{1,7},
Zhijia Tian\altaffilmark{1,7},
Huawei Zhang\altaffilmark{1},
Gaochao Liu\altaffilmark{1},
Zihuang Cao\altaffilmark{5},
Yong Zhang\altaffilmark{6},
Yonghui Hou\altaffilmark{6},
Yuefei Wang\altaffilmark{6}}

\altaffiltext{1}{Department of Astronomy, Peking University, Beijing 100871, P.\,R.\,China; 
bchen@pku.edu.cn; x.liu@pku.edu.cn}
\altaffiltext{2}{Kavli Institute for Astronomy and Astrophysics,
Peking University, Beijing 100871, P.\,R.\,China}
\altaffiltext{3}{National Astronomy Observatories, 
Chinese Academy of Sciences, Beijing 100012, P.\,R.\,China}
\altaffiltext{4}{Department of Astronomy, Beijing Normal University, Beijing 100875, P.\,R.\,China}
\altaffiltext{5}{Key Laboratory of Optical Astronomy, National Astronomical Observatories, 
Chinese Academy of Sciences, Beijing 100012, P.\,R.\,China}
\altaffiltext{6}{Nanjing Institute of Astronomical Optics \& Technology, National Astronomical Observatories, 
Chinese Academy of Sciences, Nanjing 210042, P.\,R.\,China}
\altaffiltext{7}{LAMOST Fellow}

\begin{abstract}
We select from Paper~I a sample of 306 massive star clusters observed 
with the Large Sky Area Multi-Object Fibre Spectroscopic 
Telescope (LAMOST) in the vicinity fields of M31 and M33 and
determine their metallicities, ages and masses.
Metallicities and ages are estimated by 
fitting the observed integrated spectra with stellar synthesis population (SSP) models 
with a pixel-to-pixel spectral fitting technique.
Ages for most young clusters are also derived by fitting the multi-band 
photometric measurements with model spectral energy distributions (SEDs). 
The estimated cluster ages
span a wide range, from several million years to the age of the universe.
The numbers of clusters younger and older than 1\,Gyr are {respectively 46 and 260}.
With ages and metallicities determined, 
cluster masses are then estimated by comparing the  multi-band  photometric measurements  
with SSP model SEDs. The derived masses range from $\sim 10^{3}$ to 
$\sim 10^7$\,$M_{\odot}$, peaking at $\sim 10^{4.3}$
and $\sim 10^{5.7}$\,$M_{\odot}$ for young ($< 1$\,Gyr) and 
old  ($>1$\,Gyr) clusters, respectively. 
Our estimated metallicities, ages and masses 
are in good agreement with available literature  values.
Old clusters richer than [Fe/H] $\sim -0.7$\,dex have a wide range of ages. 
Those poorer than [Fe/H] $\sim -0.7$\,dex seem to be composed of two groups, as previously 
found for Galactic GCs -- one of the oldest ages with all values of metallicity down 
to $\sim -2$\,dex and another with metallicity increasing with decreasing 
age. The old clusters in the inner disk of M\,31 (0 -- 30\,kpc) show a
clear metallicity gradient measured at $-0.038\pm0.023$ dex/kpc. 
\end{abstract}

\maketitle

\section{Introduction}

Star clusters have two distinct populations, globular  
and open clusters (GCs and OCs). GCs are old, massive, 
luminous, centrally concentrated systems. 
OCs are in general much less massive than GCs. 
They are faint, diffuse and often embedded in or 
associated with molecular clouds in the galactic disk.
However, a third population of star clusters 
the so-called young massive clusters (YMCs),
have recently been observed in many galaxies, including 
M31 \citep[and reference therein]{Fusi2005, Barmby2009, Caldwell2009,Perina2009, Kang2012, Wang2012}. 
They exhibit hybrid properties of both OCs and GCs. 
They are more massive ($\sim 10^4\,M_{\odot}$) 
than OCs while much younger ($<$1\,Gyr) than GCs.  
OCs are often faint and embedded in the dusty disk, making it  
difficult to observe and study them in a distant galaxy such as M31. 
In the current work, we focus on 
massive clusters, i.e. YMCs and GCs, in M\,31. Being luminous  objects, 
spectroscopy is feasible even with a medium-size telescope.

Many studies on the identification, 
classification, and analysis of massive clusters in M31 have been undertaken 
and published for  the past decades 
(e.g. \citealt{Barmby2000,Galleti2006, Kim2007, Caldwell2009, Perina2010}). 
\citet{Barmby2000} present $UBVRI$ and $JHK$ photometry of 435 
clusters and candidates in M31. \citet{Galleti2004} identify 
693 clusters and candidates from the Two Micron All Sky Survey (2MASS) 
database and provide an extensive Revised Bologna 
Catalogue (RBC) \footnote{ http://www.bo.astro.it/M31/}, 
including many multi-band optical data compiled from  
previous catalogs. RBC is frequently  updated, the latest is  Version 5 released in August, 2012. 
It serves as a main repository of information of  star clusters in M31. 
Starting from RBC, \citet{Caldwell2009} present a new catalog containing 670 likely star clusters. 
Most of those are confirmed ones based on high-quality spectroscopy 
with the Hectospec spectrograph mounted on the 6.5\,m MMT  telescope. 
In addition, \citet{Caldwell2009} present  ages and reddening values of 140 young clusters by 
comparing the observed and model spectra.
Based on the classification of \citet{Caldwell2009}, 
\citet{Caldwell2011} provide metallicities and ages of 367 old clusters 
derived from  the high-quality spectra. {Most recently, \citet{Caldwell2016} 
have refined a few metallicities from \citet{Caldwell2011} and 
added a small number of new observations of previously known clusters to the collection}.
\citet{Peacock2010} present an updated catalog containing 
newly collected  $ugriz$ optical photometry based on images from the 
Sloan Digital Sky Survey (SDSS; \citealt{York2000}) and 
near infrared (IR) $K$-band photometry from the
Wide Field CAMera (WFCAM) survey with the UK Infrared Telescope  (UKIRT). The catalog includes 
homogeneous photometry of 572 clusters and 373 candidates. 

\citet[hereafter Paper I]{Chen2015} present a catalog of 908 objects observed 
with the Large Sky Area Multi-Object Fiber Spectroscopic Telescope (LAMOST; 
\citealt{Cui2010, Cui2012}) in the vicinity fields of M31 and M33, 
targeted as star clusters and candidates. Most of the targets were 
selected from RBC and the SDSS catalogs of extended sources.
From the early phase of LAMOST operation, 
as parts of the LAMOST Spectroscopic Survey of the Galactic Anti-centre 
(LSS-GAC; \citealt{Liu2014, Liu2015,Yuan2015}), LAMOST has been  used to carry 
out a systematic spectroscopic survey of objects  of special interest in the vicinity 
fields of M31 and M33, including the 
star clusters  and candidates, planetary nebulae (PNe), \HII~regions, 
supergiants in M31 and M33, as well as background quasars.
The catalog presented in Paper I is based on LAMOST spectroscopic
observations of 306 star clusters and 49 candidates.
Among them, 9 clusters and 23 candidates are newly discovered.
Star clusters are formed during major star-forming episodes of a galaxy. 
The ages, metallicities and masses of individual clusters, 
in a given galaxy, trace formation and evolution history of the host galaxy. In this work, 
we focus on the estimation of metallicities, ages and masses 
of a sample of confirmed massive star clusters cataloged in Paper I. 

A variety of  methods have been  devised and applied to determine the ages and metallicities
of  star clusters in M31. We distinguish three approaches in this paper: (i) Full spectral fitting;
(ii)  Lick/IDS  absorption line indices; and (iii) Spectral energy distribution (SED) fitting.
All approaches rely  on accurate modelling of 
simple stellar populations (SSP; \citealt{Renzini1988,Bruzual2003,Kotulla2009,Vazdekis2010}
and references therein).

The full spectral fitting compares the observed integrated spectra of 
with SSP model spectra on a 
pixel-by-pixel basis, in order to derive the cluster ages and metallicities. 
The technique is an improvement compared to earlier methods and 
has recently been extensively discussed by, for example, 
\citet{Koleva2008} and \citet{Cid2010}.
Two full spectral fitting codes, STARLIGHT \citep{Cid2005} and ULySS \citep{Koleva2009}
have been developed and widely applied. \citet{Dias2010} derive ages and 
metallicities of 14 Small Magellanic Cloud (SMC)
star clusters using both codes coupled with three sets of  SSP 
model spectra from the literature. They show that the choice of code
has a larger  impact on the results than the choice of models.
\citet{Cezario2013} obtain ages and metallicities of 38 M31  and 41 Galactic GCs
using ULySS and SSP models of \citet{Vazdekis2010}. Their results are 
in good agreements with previous work.

Lick indices are easy to measure with low-resolution spectroscopy 
and even narrow-band photometry, and thus have been widely used derive properties 
 of star clusters from their integrated spectra for more than a decade (e.g. 
\citealt{Worthey1994}). The indices focus on strengths (equivalent widths) of specific features, 
and are therefore insensitive to uncertainties in  flux calibration.
\citet{Graves2008}  present a publicly available code called \eza~for determining the mean, 
light-weighted ages and abundances of Fe, Mg, C, N, and Ca of stellar 
populations from the Lick indices measured in their integrated spectra.
Calibrated with  Galactic GCs and a number of super-solar metallicity 
SSP models, \citet{Galleti2009} estimate  from  
Lick indices metallicities for  GCs in M31. 
Metallicities of GCs in M31 are also estimated by
\citet{Caldwell2011} from  Lick indices calibrated with metallicities of Galactic GCs. 
In addition, they have also determined ages  of GCs of  [Fe/H] $\ge -$1\,dex 
using program \eza. Considering that the indices are calibrated with Galactic 
GCs and the limitation of SSP models available in \eza~ 
\citep{Schiavon2007}, all those studies  are only applicable to old clusters in M31.

The SED fitting is sensitive to the general shape of continuum 
over a broad wavelength range (e.g. \citealt{Wang2010,Fan2010}).
The  method is more susceptible to the age-metallicity degeneracy
 in the sense that young metal-rich populations are 
photometrically indistinguishable from older metal-poor ones.  
The degeneracy is severe if only optical photometry is available 
\citep{Worthey1994,Arimoto1996,Kaviraj2007}. 
\citet{Anders2004} have studied star clusters in NGC 1569 
using multi-band Hubble Space Telescope 
(HST) photometry. They strongly recommend 
to include  near-infrared (NIR) photometry to break the age-metallicity degeneracy for 
young clusters (see also \citealt{deJong1996}). They 
show that the access to at least one NIR passband can 
significantly improve the results and obtain tight constrains on metallicity. 
SEDs of old clusters are usually hard to distinguish amongst themselves, 
thus the SED fitting method is usually only applicable to 
derive ages of young clusters (e.g. \citealt{Kang2012, Wang2012}).

In the current work all three methods 
are applied  to obtain robust estimates of metallicity and age of massive star cluster sample
in M31, as well as one cluster in M33, that have been homogeneously observed with LAMOST.
Masses of the clusters  are then  derived from the photometric data 
and values of mass-to-luminosity ratio $M/L_V$ 
values from SSP models. \citet{Caldwell2011}, in their mass  estimates of 
sample star clusters, assume $M/L_V$ = 2. 
However, \citet{Strader2011} show that M31 star 
clusters have a range of $M/L_V$ ratio  between 0.27 and 4.05. \citet{Ma2015} 
estimate cluster masses by comparing the photometry given by 
SSP models, adopting ages and metallicities given by  
\citet{Caldwell2011}. In the current work, cluster masses are also estimated by 
comparing the multi-band photometry with SSP models,
but using ages and metallicities determined in a 
homogeneous way presented in the same piece of work.

The paper proceeds as follows. In \S{2} we describe the 
data of our star cluster sample, including
the LAMOST spectroscopic observations  and the SDSS photometric data.
Determinations of cluster metallicities are presented in \S{3}. 
We calculate the cluster ages in \S{4} and masses in \S{5}. 
The  results are discussed in \S{6}. Finally we summarize in \S{7}.

\section{Data}

\subsection{Sample}

\begin{figure}
  \centering
  \includegraphics[width=0.48\textwidth]{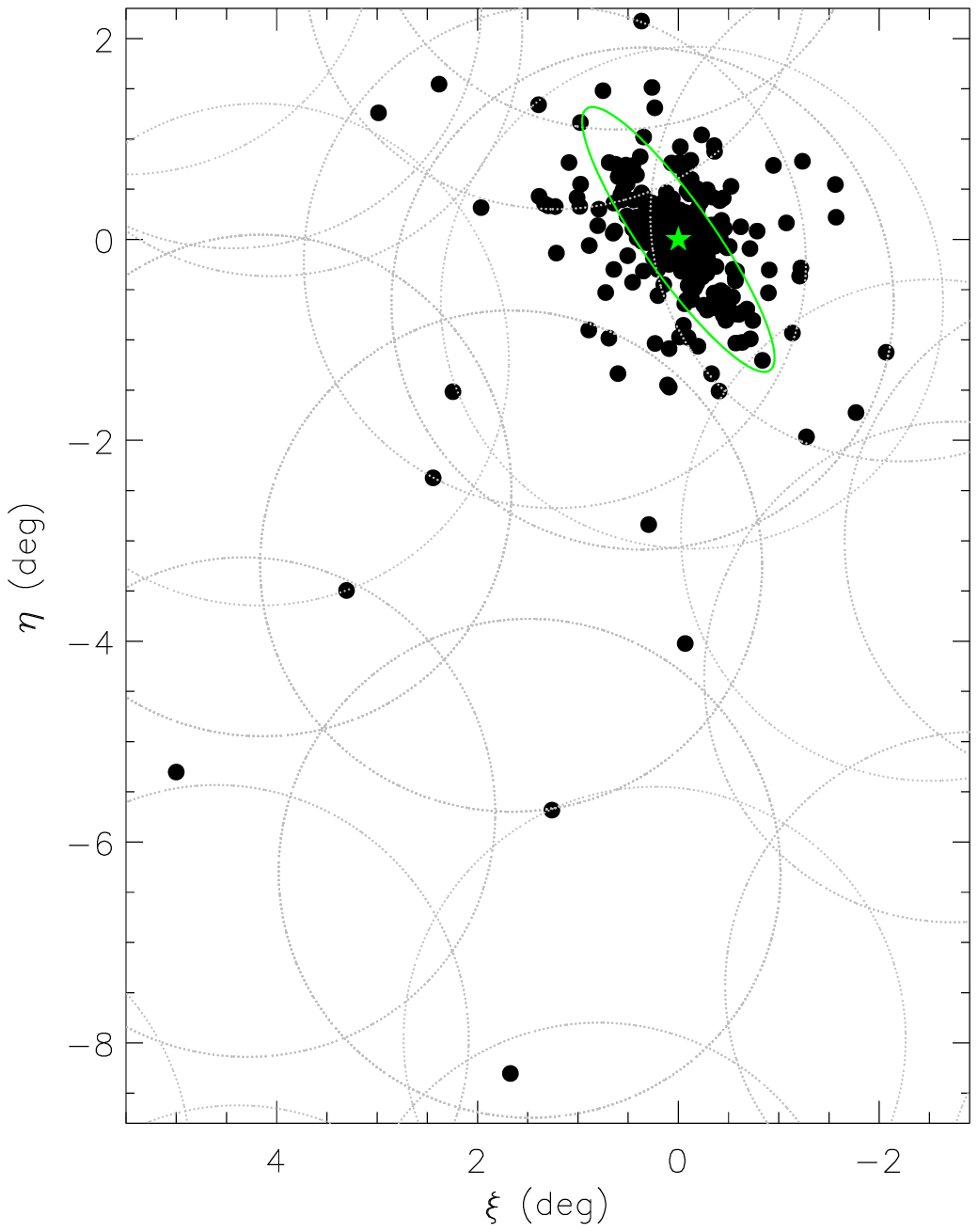}
  \caption{Spatial distribution of our sample of star clusters in M31.
{The large black circles represent LAMOST spectroscopic plates} 
 observed since June 2014.
The green star marks the central positions of M31. 
The green ellipse represents the optical disk of M31 of 
radius $R_{25}\,=\,95'.3$ \citep{deVau1991}, with
an inclination angle $i$ = 77$\degr$ 
and a position angle $P.A.$ = 38\degr \citep{Kent1989}. 
 Also included in the sample is LAMOST-2, likely to be a cluster 
 associated with M\,33. It however falls outside the field plotted here.  }
  \label{spat}
\end{figure}

Massive clusters in our sample are all selected from the catalog presented 
in Paper I, including 5 newly discovered clusters 
selected with the  SDSS photometry, 3 newly
confirmed and 298 previously known clusters from RBC. Paper~I lists 
296 known clusters from RBC. Since then, another two objects,
B341 and B207, have also been observed with LAMOST, 
and they are included in the current analysis.
The current sample do not include 
those listed in Paper~I but selected from \citet{Johnson2012}  since most of them are 
young but not so massive. All objects are observed with LAMOST between 
September, 2011 and  June, 2014. Table~1 lists the name, position and radial velocity
of all sample clusters analyzed in the current work. 
Their spatial distribution in the $\xi~-~\eta$ plane  is shown in Fig.~\ref{spat}.
Here $\xi$ and $\eta$ are respectively Right Ascension and 
Declination offsets relative to the optical centre of M31 
(RA: 00$^{\rm h}$42$^{\rm m}$44$^{\rm s}$.30; 
Dec:+41\degr16$^{\prime}$09$^{\prime\prime}$.0, 
from \citealt{Huchra1991} and \citealt{Perrett2002}).    
The clusters distributes between the thin disk and the outer halo of M31. One cluster 
identified with LAMOST, LAMOST-2, falls close to  
M33 and has a radial velocity similar to 
that of M33. It is likely a halo cluster of M33 and is included in the current analysis.

\subsection{LAMOST spectra}

\begin{figure*}
  \centering
  \includegraphics[width=1.0\textwidth]{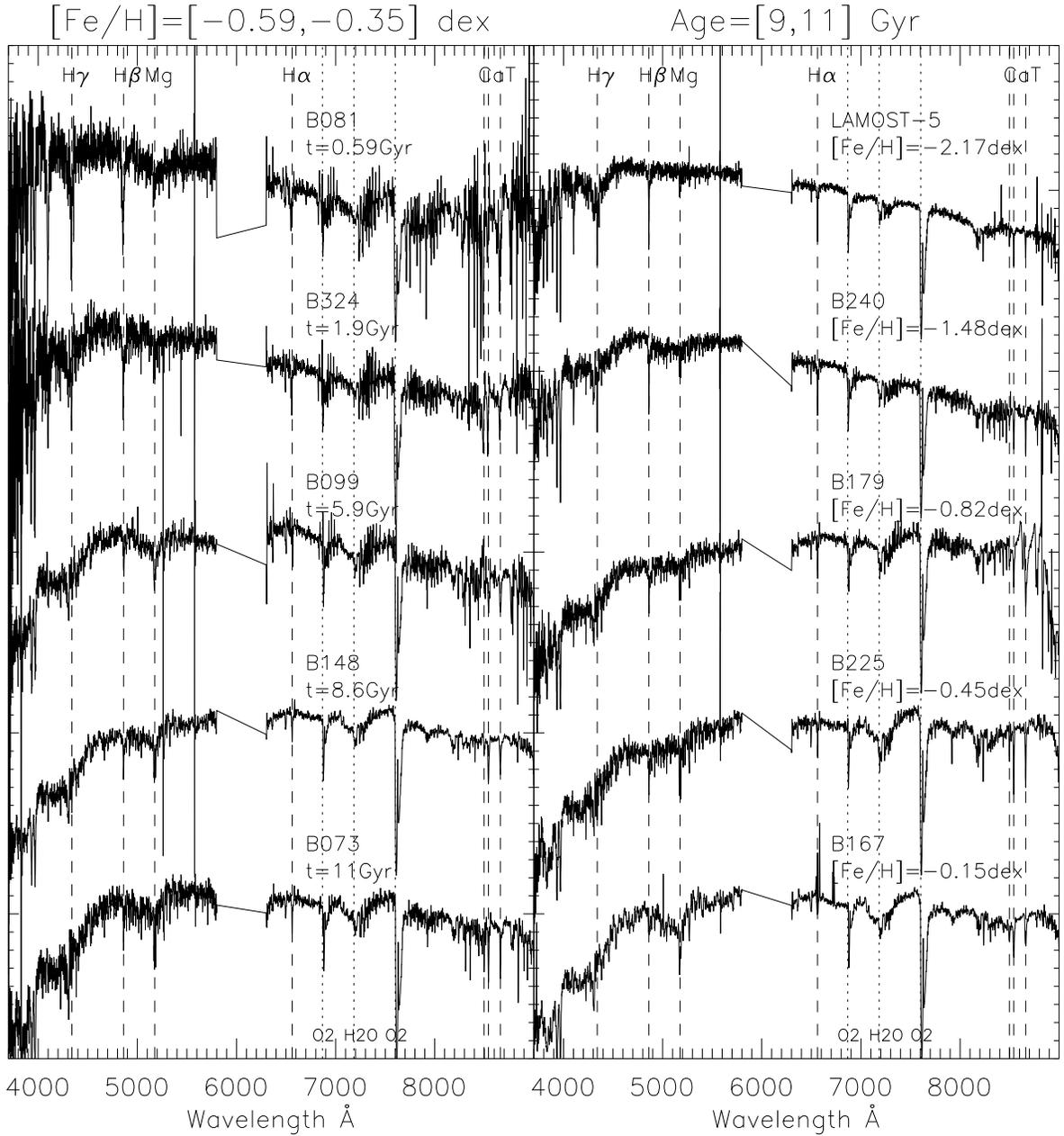}
  \caption{Example LAMOST spectra of star clusters in our sample. 
  {Note that we have removed the spectra in wavelength 
  range 5800 -- 6300 \AA~ where the blue- and red-arm spectra are connected. }
  {\em Left-hand panel}: Spectra of clusters of metallicities between 
  $-0.59$ and $-0.35$\,dex but of different ages that
  increase from top to bottom. The source name and age (derived 
  from full spectral fitting with the models of Vazdekis et al.) 
  are marked for each spectrum. 
  {\em Right-hand panel}: 
  Spectra of clusters of  ages between 9 and 11\,Gyr but of  different metallicities that  
  increase from top to bottom. The source name and metallicity (derived from full spectral 
  fitting with the models of Vazdekis et al.) are marked for each spectrum. 
  The wavelengths of all the spectra have been shifted to the rest frame
  and the flux levels shifted by arbitrary amounts.
  The vertical dashed lines mark the positions of 
  absorptions lines H$\gamma$, H$\beta$, Mg, H$\alpha$,
  and Ca~{\sc ii} triplet, respectively. {The vertical dotted lines 
  mark the positions of strong telluric lines.}}
  \label{spectra}
\end{figure*}

The LAMOST M31/M33 survey is a component of 
LSS-GAC \citep{Liu2014,Liu2015,Yuan2015}. The spectra  
were collected with nine unique but overlapping
spectroscopic plates of 5$^\circ$ in diameter (see Fig.~\ref{spat}). 
All plates were observed under  dark or 
grey lunar conditions. Typically 2--3 exposures were made for each 
plate, with typical integration time per exposure between, depending on the observing conditions, 
600--1200\,s, 1200--1800\,s and 1800--2400\,s for bright (B),  median (M) and 
faint (F) plates, respectively. Some test nights reserved for  monitoring the telescope performance 
were also used to observe M31 and M33 plates. For most observations, the seeing varied between 3 --
4\,arcsec, with a typical value of about 3.5\,arcsec \citep{Yuan2015}. 

The LAMOST spectra cover the wavelength range  
3700--9000\,\AA at a  resolving 
power of $R \sim 1800$. 
Details about the  observations and data reduction can be found  in Paper I. 
The median signal-to-noise ratio (SNR) per pixel at 4750 and 7450\,\AA~ of  
spectra of all clusters in the current 
sample are respectively 14 and 37. Essentially all spectra have
\bsnr$> 5$ except for spectra of  18 clusters. 
The latter have $\rsnr > 10$.
The spectra were first processed with the LAMOST 2-D Pipeline Version 2.6 
\citep{Luo2015}. Flux calibration was carried out using the algorithm of 
\citet{Xiang2015a}. For each plate, a number of sky fibres 
($\sim$20 for each of the sixteen spectrographs)  
were assigned, targeting blank sky areas. 
For the central area of M31 where the surface brightness from the host galaxy is high compared  
to the clusters targeted, the nearby sky fibres were used to subtract  the 
local background of  for the targets. A similar approach was adopted by 
\citet{Caldwell2009}. Partly benefited from this, 
radial velocities derived from our observations for clusters falling 
close to the central bright bulge are in close agreement with 
those obtained by \citet{Caldwell2009}.
For objects observed multiple times 
with  different plates, spectra with lower \bsnr~  were  scaled with a  low-order polynomial 
to match the continuum level of spectrum of the highest and then 
all combined together, weighted by the spectral invariances.
Radial velocities of the clusters were derived 
 by matching the observed spectra with SSP models. 
 Example spectra of several star clusters of different metallicities 
and ages (see \S\S{3} and 4) are presented in Fig.~\ref{spectra}.
Note that the blue- and red-arm spectra 
were processed separately with the 2-D Pipeline and joined together after the flux 
calibration \citep{Xiang2015a}. No scaling nor shifting was performed in cases 
where the blue- and red-arm spectra did not have the same flux level in the 
overlapping wavelength region, as it was unclear whether the misalignment was caused by 
poor flat-fielding/flux-calibration  or sky subtraction, or 
a combination of both. That is why spectra of some clusters show abnormal 
artefacts in wavelength range 5800 -- 6300\,\AA. 
We have removed this part of spectra in Fig.~\ref{spectra} {for better illustration.} Finally 
the bright sky emission line  at 5578\,\AA~were not properly removed in spectra of some 
objects. 
 
\subsection{Photometric Data}

\citet{Peacock2010} retrieved  images of M31 star clusters and candidates 
from the SDSS archive and extracted $ugriz$ aperture photometric magnitudes those objects 
using SE$\rm {\scriptstyle XTRACTOR}$.
They present a catalog containing  
homogeneous $ugriz$ photometry of 572 star clusters and 373 candidates.
Amongst them, 299 clusters are in  our sample. Of those, 
280, 289, 289, 287 and 285 are detected in $u,~g,~r,~i$ and $z$ bands, respectively.
Six objects, including the five newly discovered clusters 
reported in Paper I (LAMOST-1--5) and
MGC1 \citep{Huxor2008} are not found in the Peacock et al. catalog but 
have been detected by SDSS. For the six objects, we  adopt the Petrosian magnitudes
available from the SDSS archive. 

Also listed in Table~1 are $ugriz$ photometric magnitudes along  with  
reddening values. The reddening values are taken from \citet{Kang2012}, 
compiled from three sources: (1) Determinations available
from the literature  (\citealt{Barmby2000,Fan2008,Caldwell2009,Caldwell2011}). 
If more than one determinations are available, the average value is adopted;
(2) For clusters that have no reddening values determinations in the literature, the
median value of clusters within 2\,kpc radius of the cluster of concern is adopted; and 
(3) For star clusters that fall beyond a galactocentric distance of 22\,kpc, 
a foreground reddening value of $E (B - V )$ = 0.13\,mag is adopted. 
Some of our sample clusters are not available in the compilation of Kang et al.
For those objects, we calculate their reddening values following approaches (2) and (3) above. 

\begin{table*}
 \centering
  \caption{Optical photometry and reddening of massive star clusters observed with LAMOST}
  \begin{tabular}{lrrrcccccc}
  \hline
  \hline
Name & R.A. & Decl. & $V_r$  & $u$ & $g$ & $r$ & $i$ & $z$ & $E(B-V)$  \\
          & (deg) & (deg) & (\kms)  & (mag) &  (mag) & (mag) &  (mag) & (mag)  & (mag) \\ 
 \hline
 B001-G039 &     9.96253 &    40.96963 &  -222$\pm$  150   & 19.39$\pm$ 0.12   & 17.58$\pm$ 0.07   & 16.61$\pm$ 0.07   & 16.07$\pm$ 0.07   & 15.70$\pm$ 0.07   &  0.25   \\
 B002-G043 &    10.01072 &    41.19822 &  -370$\pm$   18   & 19.15$\pm$ 0.10   & 17.86$\pm$ 0.08   & 17.34$\pm$ 0.07   & 17.06$\pm$ 0.07   & 16.90$\pm$ 0.09   &  0.01   \\
 B003-G045 &    10.03917 &    41.18478 &  -381$\pm$    9   & 19.43$\pm$ 0.12   & 17.94$\pm$ 0.09   & 17.36$\pm$ 0.07   & 16.99$\pm$ 0.07   & 16.82$\pm$ 0.09   &  0.16   \\
 B004-G050 &    10.07462 &    41.37787 &  -381$\pm$    4   & 19.06$\pm$ 0.10   & 17.40$\pm$ 0.07   & 16.64$\pm$ 0.07   & 16.28$\pm$ 0.07   & 16.05$\pm$ 0.07   &  0.13   \\
 B005-G052 &    10.08462 &    40.73287 &  -299$\pm$    2   & 17.86$\pm$ 0.08   & 16.12$\pm$ 0.07   & 15.32$\pm$ 0.06   & 14.90$\pm$ 0.06   & 14.62$\pm$ 0.06   &  0.22   \\
 B006-G058 &    10.11031 &    41.45740 &  -244$\pm$    2   & 17.69$\pm$ 0.07   & 15.92$\pm$ 0.07   & 15.16$\pm$ 0.06   & 14.79$\pm$ 0.06   & 14.53$\pm$ 0.06   &  0.11   \\
 B008-G060 &    10.12613 &    41.26907 &  -330$\pm$    4   & 19.03$\pm$ 0.10   & 17.23$\pm$ 0.08   & 16.47$\pm$ 0.07   & 16.07$\pm$ 0.07   & 15.86$\pm$ 0.07   &  0.17   \\
 B010-G062 &    10.13154 &    41.23956 &  -195$\pm$   11   & 18.46$\pm$ 0.09   & 17.04$\pm$ 0.08   & 16.38$\pm$ 0.07   & 16.02$\pm$ 0.07   & 15.80$\pm$ 0.07   &  0.20   \\
 B011-G063 &    10.13282 &    41.65474 &  -249$\pm$    8   & 18.42$\pm$ 0.08   & 17.06$\pm$ 0.07   & 16.43$\pm$ 0.07   & 16.14$\pm$ 0.07   & 15.97$\pm$ 0.07   &  0.09   \\
     B011D &    10.21508 &    40.73504 &  -538$\pm$   10   & 18.82$\pm$ 0.09   & 17.87$\pm$ 0.13   & 17.21$\pm$ 0.07   & 16.49$\pm$ 0.07   & 15.77$\pm$ 0.07   &  0.28   \\
 B012-G064 &    10.13525 &    41.36226 &  -367$\pm$    4   & 16.76$\pm$ 0.07   & 15.43$\pm$ 0.07   & 14.83$\pm$ 0.06   & 14.53$\pm$ 0.06   & 14.36$\pm$ 0.06   &  0.11   \\
 B013-G065 &    10.16022 &    41.42328 &  -422$\pm$    8   & 19.19$\pm$ 0.11   & 17.63$\pm$ 0.10   & 16.88$\pm$ 0.07   & 16.46$\pm$ 0.07   & 16.19$\pm$ 0.08   &  0.13   \\
      ... &    ... &    ... &  ...   & ...   & ...   & ...   & ...   & ...   &  ...   \\  
  KHM31-74 &    10.22055 &    40.58880 &   -55$\pm$    8   & 20.28$\pm$ 0.19   & 18.51$\pm$ 0.19   & 17.84$\pm$ 0.08   & 17.51$\pm$ 0.08   & 17.21$\pm$ 0.11   &  0.26   \\
  LAMOST-1 &    12.23263 &    35.56682 &   -55$\pm$    2   & 19.24$\pm$ 0.14   & 18.89$\pm$ 0.04   & 18.13$\pm$ 0.03   & 17.71$\pm$ 0.04   & 17.56$\pm$ 0.10   &  0.13   \\
  LAMOST-2 &    24.07521 &    30.27437 &  -175$\pm$    8   & 20.07$\pm$ 0.26   & 18.64$\pm$ 0.03   & 17.99$\pm$ 0.02   & 17.73$\pm$ 0.03   & 17.63$\pm$ 0.08   &  0.13   \\
  LAMOST-3 &    11.18990 &    43.44303 &  -424$\pm$    8   & 19.02$\pm$ 0.05   & 17.79$\pm$ 0.01   & 17.21$\pm$ 0.01   & 16.96$\pm$ 0.01   & 16.74$\pm$ 0.03   &  0.13   \\
  LAMOST-4 &     9.03580 &    39.29165 &  -230$\pm$   16   & 19.19$\pm$ 0.05   & 17.88$\pm$ 0.01   & 17.31$\pm$ 0.01   & 17.05$\pm$ 0.01   & 16.98$\pm$ 0.03   &  0.13   \\
  LAMOST-5 &    14.73496 &    42.46061 &  -144$\pm$    4   & 17.66$\pm$ 0.02   & 16.43$\pm$ 0.00   & 15.88$\pm$ 0.00   & 15.61$\pm$ 0.00   & 15.36$\pm$ 0.01   &  0.13   \\
      M086 &    11.36870 &    41.82488 &   -56$\pm$   10   & 19.70$\pm$ 0.14   & 18.29$\pm$ 0.09   & 17.90$\pm$ 0.08   & 17.88$\pm$ 0.10   & 17.70$\pm$ 0.14   &  0.18   \\
      ... &    ... &    ... &  ...   & ...   & ...   & ...   & ...   & ...   &  ...   \\ 
    \hline
  \end{tabular}\\
\begin{flushleft}
Note: This is a sample of the full table, which is available in its entirety in the electronic version of this 
article.
\end{flushleft}
\end{table*}

\section{Metallicities}

In this Section, we discuss measurement of cluster metallicities, 
characterised by [Fe/H]. To proceed, we first apply the full spectral fitting method.
The SSP models that we adopt for the full spectral fitting span a wide range of age,  
from young to very old. Thus the fitting is applied 
to both young and old clusters in the sample. In addition, we have also measured the 
Lick indices from the LAMOST spectra. 
To obtain estimates of [Fe/H], the Lick indices 
are calibrated by integrated spectral indices of Milky Way (MW)
GCs whose metallicities have previously been measured by
high resolution spectroscopy and elemental abundance analysis  
of bona fide cluster member stars. The 
calibration implicitly assumes that all M31 and MW clusters have the same age.
We have also determined the metallicities using the \eza~code.
The code compares the measured Lick indices to those from the SSP models of 
 \citet{Schiavon2007}. 
The models have ages ranging from 1 to 15\,Gyr 
and [Fe/H] ranging from $-$1.3 to $+$0.3\,dex.
Thus in both cases, the method 
 bases on Lick indices is only applicable to old (age $> 1$\,Gyr) clusters in our sample. 

\subsection{Metallicities from full spectral fitting}

\begin{figure*}
  \centering
  \includegraphics[width=0.49\textwidth]{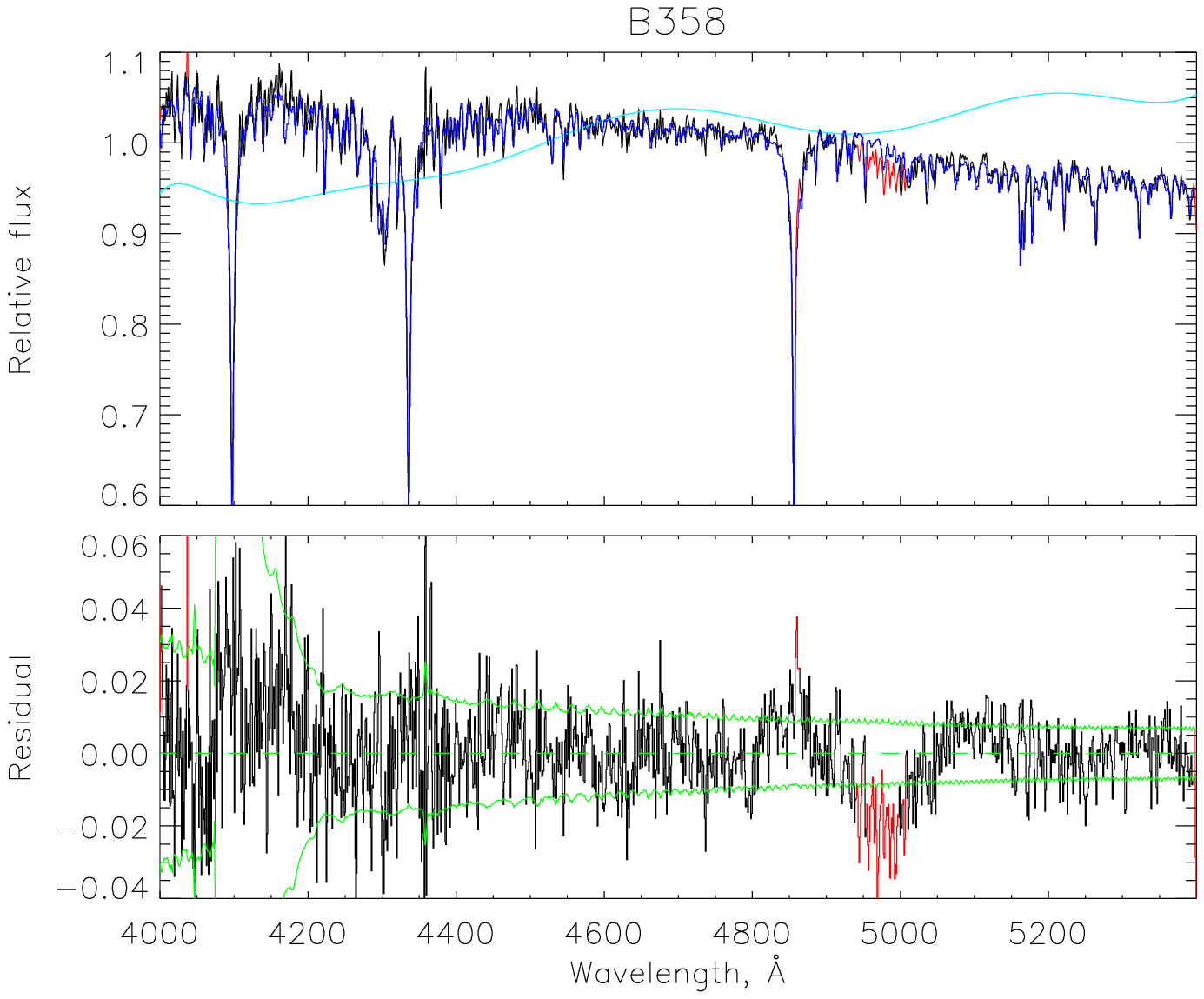}  
  \includegraphics[width=0.49\textwidth]{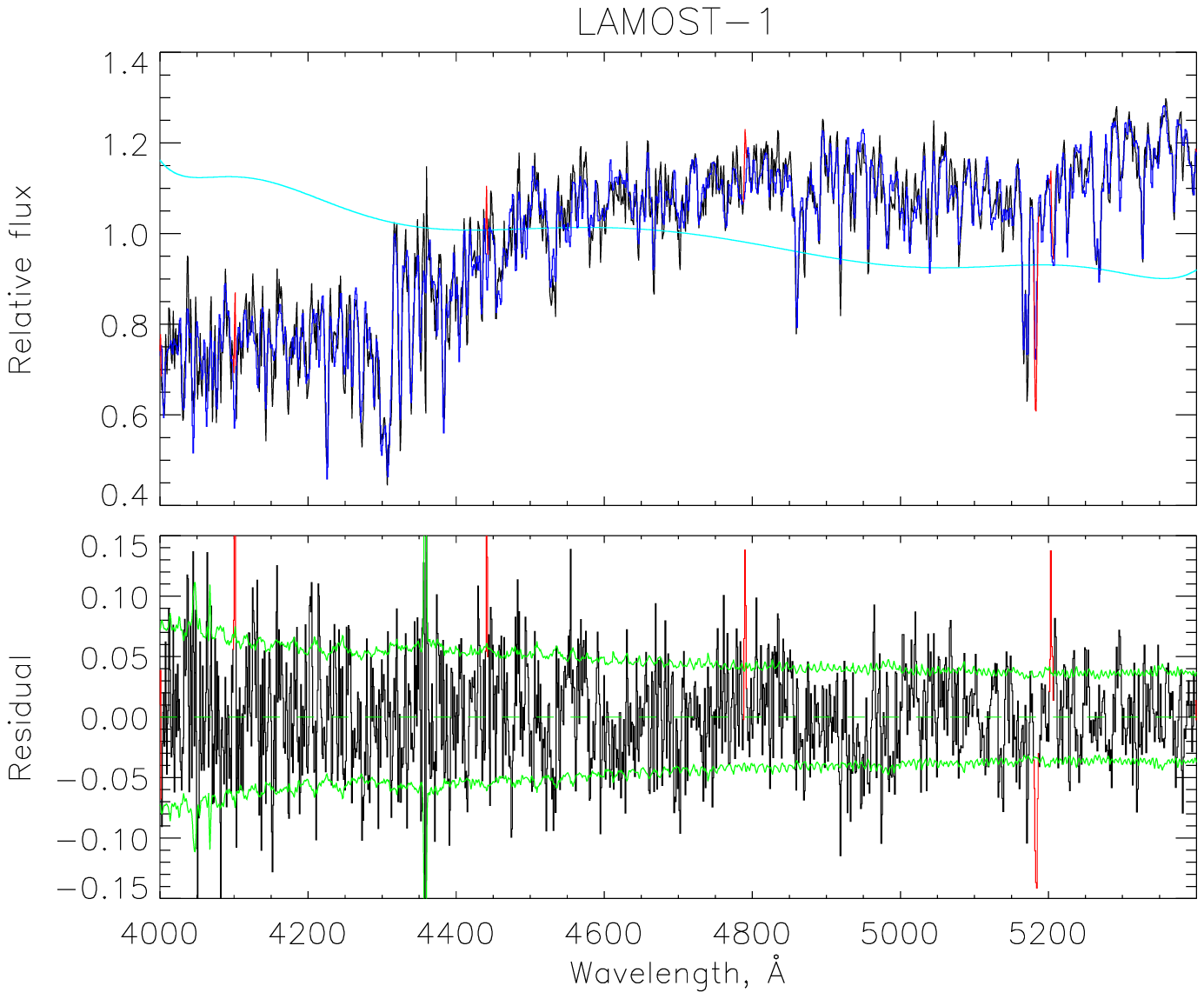}
  \caption{Example full spectral fitting of clusters B358 
  (left-hand panels) and LAMOST-1 (right-hand panels). 
  The top panels show the observed spectra in black and the best fit 
  model spectra from \citet{Vazdekis2010} in blue. The cyan lines delineate  the 
  multiplicative polynomials. The bottom panels are the fractional residuals of the best fits, where the 
  dashed and solid lines in green denote respectively zero and  
  the 1-$\sigma$ deviations, with the latter calculated from 
  the invariances of the input (observed) spectra. }
  \label{ulyss}
\end{figure*}

\begin{figure}
  \centering
  \includegraphics[width=0.48\textwidth]{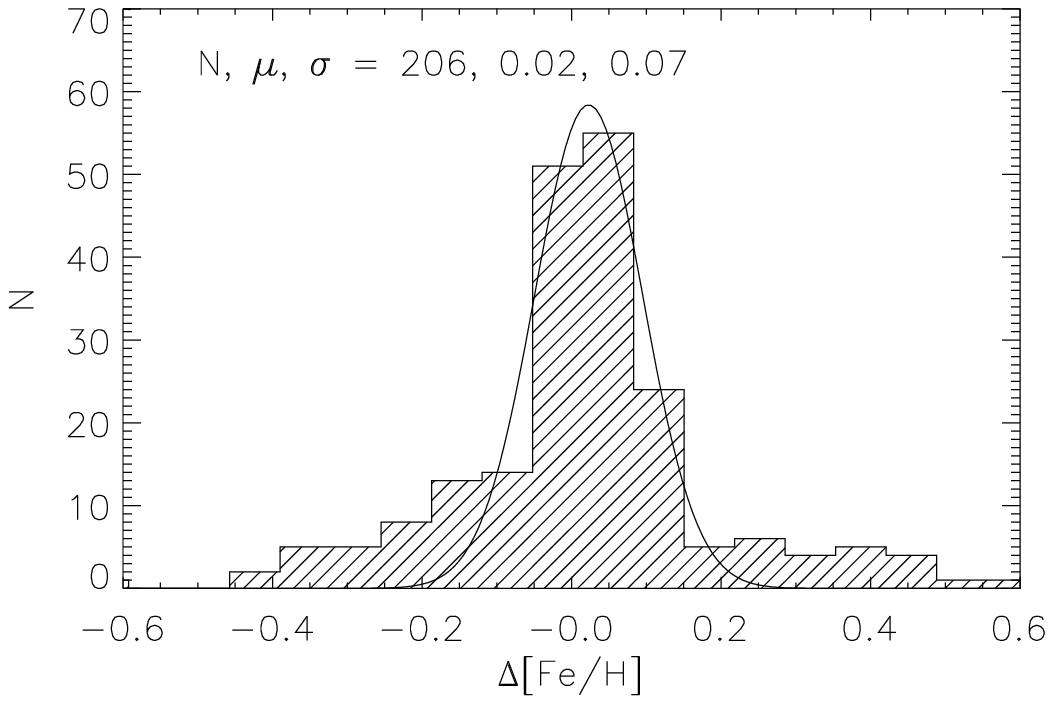}
  \caption{Histogram of differences of metallicities deduced from 
  duplicate observations of the same clusters. Overplotted is a Gaussian fit to the 
  distribution, with the number of clusters with duplicate observations, and 
  the mean and dispersion of the Gaussian marked. }
  \label{intfeh}
\end{figure}

Comparison between the observed and template spectra has  been widely used to 
derive stellar atmospheric parameters from large number of medium- to low-resolution 
spectra (e.g. LSP3, \citealt{Xiang2015b}; LASP, \citealt{Luo2015}; and references therein). 
More recently, the technique has become increasingly common for studying the integrated 
spectra of star clusters \citep{Dias2010, Cezario2013}. The method makes 
use of all information encoded in a spectrum, making it possible to perform 
analysis even at lower SNRs. In some {implementations}, the method  is
also insensitive to uncertainties in extinction correction or flux calibration.
We apply this method to all star clusters in our sample.
A  comprehensive set of templates covering wide
parameter space is of fundamental importance for the full spectral fitting. 
SSP models constructed from both empirical and synthetic spectral libraries have been 
published in the literature. In the current work, we adopt the SSP models presented by 
\citet{Vazdekis2010}\footnote{http://miles.iac.es/pages/ssp-models.php}
and \citet{Le2004}\footnote{http://www2.iap.fr/pegase/pegasehr/}. Both 
models are based on empirical spectral libraries.  
The advantage of empirical libraries is that they consist of spectra of {real} stars. 
The public code ULySS \citep{Koleva2009}\footnote{http://ulyss.univ-lyon1.fr/} 
is used for the fitting. Below we describe the two sets of SSP models and 
the ULySS code briefly.

\citet{Vazdekis2010} present SSP models covering optical wavelength range 
3540.5$-$7409.6\,\AA~ at a  nominal 
resolution of  full width at half-maximum (FWHM) of 2.3\,\AA. ~
The  models are based on the empirical stellar spectral library, 
Medium resolution INT Library of Empirical Spectra
(MILES, \citealt{Sanchez2006, Cenarro2007}).
The MILES template stars were selected to optimise the stellar parameter 
coverage required for population synthesis modelling. 
The stellar spectra were carefully flux-calibrated to an accuracy of 
a few per cent. \citet{Xiang2015b} use the
MILES library  for  LSS-GAC stellar parameter determinations.
The SSP models of \citet{Vazdekis2010} are 
using the solar-scaled theoretical isochrones of \citet{Girardi2000}.
Parameters of the isochrones, $T_{\rm eff}$, log$\,g$ and [Fe/H], are transformed 
to the observational space by means of empirical relations between colors and stellar 
parameters. Seven initial mass functions (IMFs) are used. In the current work, we adopt models 
calculated with the  \citet{Salpeter1955} IMF. The models cover
ages of $10^8 < t< 1.5 \times 10^{10}$\,Gyr for a 
metallicity range  from super-solar [M/H] = $+$0.22\,dex (Z=0.03) 
to [M/H] = $-$2.32\,dex (Z=0.0004).  Note that in the current, [M/H] and [Fe/H] are 
treated as interchangeable. 

\citet{Le2004} present SSP models PEGASE-HR covering wavelength range 
4000$-$6800\,\AA at a resolution of  FWHM $\sim$ 0.55\,\AA. 
The models are constructed using the empirical spectral library 
ELODIE \citep{Prugniel2001,Prugniel2007}.
ELODIE spectra were collected  using an 
echelle spectrograph with a very high spectral resolution ($R ~\sim$ 42, 000).
\citet{Luo2015} adopted  the  library for LAMOST stellar parameters determinations.
\citet{Le2004} use the spectrophotometric model of galaxy evolution 
PEGASE.2 \citep{Fioc1997}  to construct the SSP models.
Two different initial mass functions (IMFs) are used. 
In the current work, we adopt PEGASE-HR models calculated with  
the \citet{Salpeter1955} IMF. 
The models covers ages $10^7 < t< 1.5 \times 10^{10}$\,Gyr for a 
 metallicity range from [Fe/H] = $-$2.0\,dex (Z=0.0004) 
to [Fe/H] = $+$0.4\,dex (Z=0.05). 

The full spectral fitting code ULySS \citep{Koleva2009} is adopted
to fit the LAMOST  spectra of M31 star clusters with SSP models.
ULySS is an open-source package that fits spectroscopic observations 
against a model through a non-linear least-squares 
minimisation. The model is expressed as a linear combination of non-linear 
components convolved with a line-of-sight velocity distribution and multiplied 
by a polynomial continuum. ULySS performs spectral fitting in two astrophysical 
contexts: the determination of stellar atmospheric parameters and 
the study of stellar populations of galaxies and star clusters.
In the current work, we use the set of routines for stellar population study. The code is 
written in the GDL/IDL language.
In  contrast to other stellar analysis programs that require the observed spectrum to be 
normalized to a pseudo-continuum as a prerequisite, ULySS determines the normalization 
in the fitting process, by introducing  a multiplicative polynomial for the scaling of model spectra. 
Therefore, ULySS is relatively insensitive to uncertainties in flux calibration, 
extinction correction, or any other sources of error that affect the 
shape of the observed spectrum. We run ULySS 
with its global minimisation option. 
The wavelength range adopted for the fitting has some impact on the results.
After some tests, we decide to use the 
wavelength range 4000$-$5400\,\AA~(similar to that used by 
of \citealt{alves2009} and \citealt{Cezario2013}) for clusters with a spectra \bsnr $>$ 5 and 
6100$-$6800\,\AA~for clusters with a \bsnr $<$ 5 but with {a \rsnr $>$ 10}.

Ages and metallicities derived with ULySS are presented in Table 3. 
For illustration, Fig.~\ref{ulyss} shows the fits of two example clusters, 
B358 a metal-poor and relatively young cluster
and LAMOST-1, a  metal-rich and very old cluster. 
The model spectra shown in the Figure are those from Vazdekis et al. 
Owing to the overlapping of Field of View (FoV) of adjacent plates, and 
some repeated observations, either because the first  observations failed to 
pass the quality control (60\% of the targets pass the 
SNR requirement, SNR(4750\AA) or SNR(7450\AA) $>10$) or 
some other reasons, over half clusters
in our sample were observed  more than once with 
LAMOST (c.f. \citealt{Liu2014, Yuan2015} and Paper I).
Those multi-epoch duplicate observations of the same targets 
provide an opportunity to test the precision of parameters 
yielded  by the full spectral fitting method. 
Fig.~\ref{intfeh} shows the distribution of differences of metallicities
deduced from the duplicate observations of the same targets. 
Only results  based on SSP models of the Vazdekis et al. 
are shown because those from the PEGASE-HR models are quite similar.
In general, the Figure shows that metallicities yielded by full spectral fitting 
have a precision better than $\sim$0.1\,dex.

\begin{figure*}
  \centering
  \includegraphics[width=0.68\textwidth]{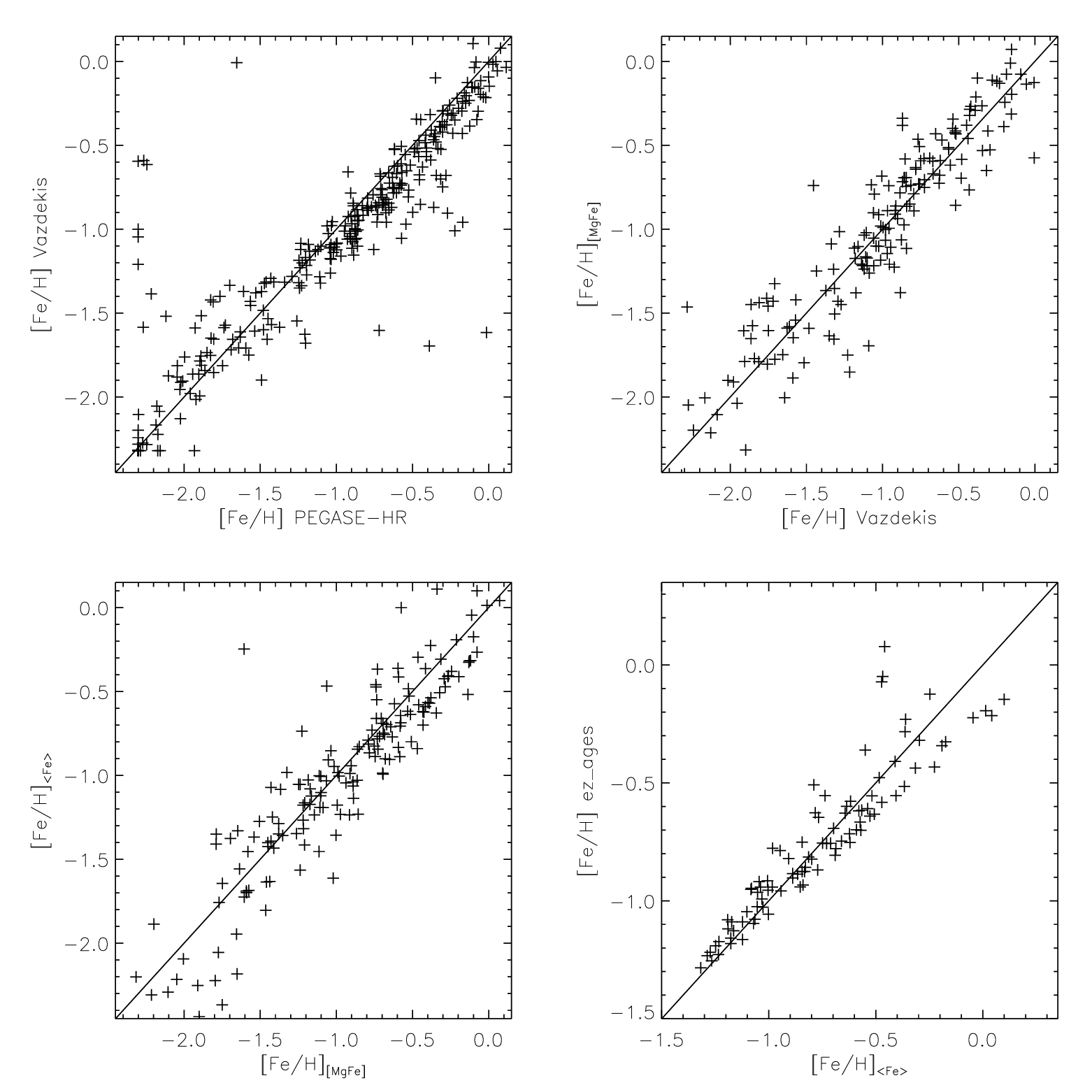}
  \caption{Comparison of metallicities derived from various methods employed in the current work. 
  The solid line in each panel denotes perfect agreement. { [Fe/H]$_{\rm PEGASE-HR}$,
[Fe/H]$_{\rm Vazdekis}$,  [Fe/H]$_{\rm [Mg/Fe]}$, [Fe/H]$_{\rm <Fe>}$
and [Fe/H]$_{\rm ez\_ages}$ are 
clusters metallicities from full spectral fitting with the  PEGASE-HR models,
from full spectral fitting with the models of  Vazdekis et al,
from the [MgFe] index using the relation of  \citet{Galleti2009},
from the \avfe~index using the relation of  \citet{Caldwell2011} and from
the \eza~package, respectively.}  }
  \label{fehlick}
\end{figure*}

Many clusters in our sample have previously been studied, 
some by fitting the color-magnitude diagrams (CMDs) with isochrones  
while others using  spectral indices. 
We will compare our results with those in the literature in \S{3.3}. 
The top left panel of Fig.~\ref{fehlick} compares 
metallicities derived with ULySS but using different SSP models. 
Overall, the agreement is good. For clusters of metallicities [Fe/H] $>-1.0$\,dex,
the values derived using the  PEGASE-HR models 
are slightly  higher ($\sim$0.07\,dex) than those from the models of 
Vazdekis et al. models. For some clusters, the 
differences are quite significant. Most of these clusters have ages smaller than 
2\,Gyr (see \S{3.3}). Excluding those outliers, the dispersion of
the differences is very small ($\sim$0.1\,dex).

\subsection{Metallicities from the  Lick Indices}

\begin{figure}
  \centering
  \includegraphics[width=0.48\textwidth]{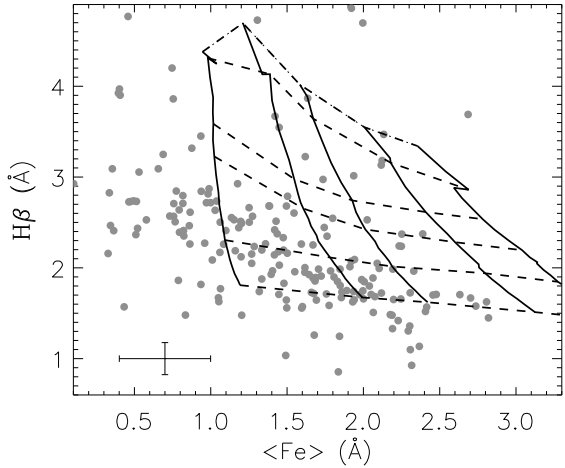}
  \caption{Loci of \avfe--H$\beta$ indices from 
  \citet{Schiavon2007}, with index measurements for M31 clusters overplotted. 
The solid  lines represent loci of different metallicities -- from left to right, 
[Fe/H] = $-$1.3, $-$0.7, $-$0.4, 0.0, and +0.2\,dex, 
  whereas dashed lines represent loci of different ages --  from bottom to top, 
$ t$ = 14, 8, 3.5, 2.5, 1.5, and 0.9\,Gyr.
{The error bar in the bottom left corner represents the average errors of measurements.}}
  \label{lickgrid}
\end{figure}

\begin{table}
 \centering
  \caption{Zero points to convert our measurements of equivalent widths 
to the Lick system from literature.}
  \begin{tabular}{lrrrrr}
  \hline
  \hline
Index & Mg$_2$ & Mg$b$ & Fe5270 & Fe5335 & H$\beta$ \\
 \hline
 Zero point$^a$ (\AA)  & 0.01  & $-$0.18  & $-$0.04 & 0.20 & 0.10 \\
 rms$^a$ (\AA)       & 0.03      & 0.56   & 0.55   & 0.59 &  0.59 \\
 Zero point$^b$ (\AA)  & 0.02  & $-$0.14  & $-$0.11 & 0.07 & 0.10 \\
 rms$^b$ (\AA)       & 0.04     & 0.46   & 0.07   & 0.46 &  0.31 \\
 \hline
\end{tabular}
\begin{flushleft}
$^a$ Zero point=$I_{\rm G09}-I_{\rm this work}$.\\
$^b$ Zero point=$I_{\rm S12}-I_{\rm this work}$.\\
\end{flushleft}
  \label{tal}
\end{table}

Lick indices have been  widely used for over a decade for 
determining ages and metallicities of star clusters. 
We calculate line indices in the Lick/IDS system 
(e.g. \citealt{Worthey1994} and references therein) 
for old clusters with LAMOST spectral \bsnr $>$ 10 in our sample.
The observed spectra are  shifted to rest frame wavelengths such that a 
 consistent set of wavelengths can be used to define and calculate line indices.
To measure the line indices, we use the code $\rm {\sc lick_ew}$ that comes with the \eza~package 
\citep{Graves2008}. The LAMOST spectra were  smoothed to match the (lower) Lick/IDS resolution 
\citep{Worthey1997}. The equivalent widths (EWs) were then measured adopting 
the passbands defined by \citet{Worthey1994} and \citet{Worthey1997}. 
Values of metallicity [Fe/H] of old clusters in our sample can then 
be derived from the line measured indices using the empirical relations.
For the latter, \citet{Galleti2009} have recently derived 
a new relation between [Fe/H] and [MgFe], 
where \mgfe$\equiv {\rm \sqrt{{\rm Mgb\cdot \langle Fe \rangle}} }$, 
and  \avfe=(Fe5270+Fe5335)/2 \citep{Gonzalez1993}. The relation is
based on well-studied Galactic GCs supplemented by theoretical models for 
$-$0.2 $\leq$ [Fe/H] $≤\leq$ $+$0.5\,dex,
\begin{eqnarray}
    {\rm[Fe/H]_{[MgFe]}} = {\rm -2.563 +1.119[MgFe] - 0.106[MgFe]^2}. 
    \nonumber
\end{eqnarray} 
For prudence,  the application of the relation should be limited to clusters 
of ages older than 7$-$8\,Gyr (see \citealt{Galleti2009}).
Alternatively, \citet{Caldwell2011} estimate metallicities of M31 GCs based on 
\avfe, again calibrated using measurements of metallicity 
[Fe/H] of Galactic GCs, 
\begin{eqnarray}
    {\rm [Fe/H]_{\langle Fe \rangle }} = - 2.23 + 0.83  {\rm \langle Fe \rangle;~ for~\langle Fe \rangle > 0.9}; \nonumber \\
    {\rm[Fe/H]_{\langle Fe \rangle }} = - 3.18 + 1.88  {\rm \langle Fe \rangle;~ for~\langle Fe \rangle 􏰏 \leq 0.9}. 
    \nonumber
\end{eqnarray}
Both calibrations  assume that GCs of M31 have similar ages,  
ignoring the fact  that clusters in each of the two galaxies actually span 
a range of ages and the age distribution can be quite different. In the 
current work, we calculate metallicities of clusters in our sample using both relations.
In doing so, the instrumental EWs are converted to the corresponding Lick/IDS indices using zero points 
derived from objects in our sample that are in  common with those in the literature 
(\citealt{Galleti2009} for the relation of \citealt{Galleti2009}; and \citealt{Schiavon2012}
for both the relations of \citealt{Caldwell2011} and that adopted in the code \eza).~We concentrate on 
a few lines indices including Mg$_2$, Mgb, Fe5270, Fe5335 and H$\beta$. The zero 
points are determined  by the averages  of differences  between our measurements and those
from the literature. They are listed in Table~2, together with the rms of the differences.

Ages and metallicities of clusters in M31 can also be determined 
from the Lick indices using code \eza,
developed by \citet{Graves2008} for automatic stellar population analysis.
\eza~is a package written in IDL that computes the mean, light-weighted age, metallicity [Fe/H], 
and elemental abundances [Mg/Fe], [C/Fe], [N/Fe], and [Ca/Fe] for unresolved stellar populations. 
This is accomplished by comparing the measured 
Lick indices with predictions of SSP models of 
\citet{Schiavon2007}, using a method described in \citet{Graves2008}. 
The method has been successfully tested 
by  applying  to Galactic GCs of  known ages and chemical composition. 
Ages and abundances are determined by using the index-index model 
grids. In the current work, we use the H$\beta$ -- \avfe~index pair. The pair 
provide good estimates of metallicities and ages of star clusters.
In additional the H$\beta$ and \avfe~ indices are relatively easy to measure and 
 are insensitive to elemental abundance variations.
Fig.~\ref{lickgrid} plots  the \avfe -- H$\beta$ grid, 
with our measurements of M31 clusters overplotted. 
{There are a relatively large fraction of objects located outside the model grid. 
Some of them are metal poor clusters.  They have blue horizontal
branch stars which may not be represented in the \citet{Schiavon2007} models. 
The young clusters are not included in the model,
either. In addition, there are some old clusters 
whose H$\beta$ line indexes are so small that they 
fall below the predicted values of the oldest 
models. Most of them are actually consistent 
with the models if we consider the index measurement errors.
A minority of them falls far away from the oldest models and that could not be 
caused by the measurement uncertainties alone.
In those cases, the discrepancies may be partly 
caused by the model zero-point uncertainties \citep{Schiavon2002} and 
the contaminations of Balmer lines from the evolved giants and/or intra-cluster 
medium \citep{Poole2010, Caldwell2011}.} 
\eza~does not deal with model extrapolation,  so 
clusters with line index measurements falling outside the model grid are excluded from 
the analysis. This includes clusters of metallicities [Fe/H]$<-$1.3 or
[Fe/H]$>+0.2$\,dex.

Fig.~\ref{fehlick} compares values of [Fe/H] estimated  from our spectra 
using various  methods described above. Overall the agreement is  good, 
with negligible  offsets and small dispersions. The rms differences 
of values  derived with  ULySS using the SSP models of Vazdekis et al.  
and those from the relation of \citet{Galleti2009} is 0.22\,dex. The corresponding value between results
derived from relation of \citet{Galleti2009} and those from relation of  \citet{Caldwell2011} 
is 0.25\,dex, and 0.13\,dex between the results  derived from the relation of \citet{Caldwell2011}  
and those yielded by code  \eza. 

\subsection{Comparison with previous [Fe/H] measurements}
\label{fehcomp}

\begin{figure}
  \centering
  \includegraphics[width=0.48\textwidth]{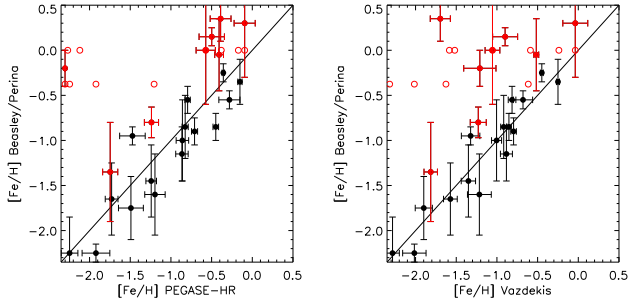}
  \caption{Comparison of our metallicities derived from 
full spectral fitting using the  PEGASE-HR models (left panel) and
models by  Vazdekis et al. (right panel) and 
  those published by Beasley et al. (2014; red filled circles) 
   and by Perina et al. (2010; red open circles).  
For completeness, old clusters from Beasley et al. are also plotted in black symbols.   
  The solid lines indicate  full agreement.}
  \label{fehyoung}
\end{figure}

Only a few studies on the metallicity of young clusters in M31 is available in the literature. 
Based on high-quality spectra, \citet{Beasley2004} obtain  metallicities for 30  
star clusters in M31. Eight of them are young clusters.
\citet{Perina2010} derive ages and metallicities of 25 young clusters by fitting the optical 
color-magnitude diagrams with theoretical isochrones using {\em HST}/WFPC2 data. 
Fig.~\ref{fehyoung} compares our metallicities derived from full spectral fitting using various
SSP models and those of  \citet{Beasley2004} and \citet{Perina2010}.
For the young clusters, except for a few outliers, our estimates, deduced using 
PEGASE-HR models or those of Vazdekis et al. regardless,  correlate well with 
those obtained by Beasley et al. (2004) but with values $\sim 0.45$\,dex systematically 
lower. Similarly, our estimates are systematically lower than those of Perina et al. (2010). 
Note that the latter have only two distinct values, either of ${\rm Fe/H]} = -0.38$\,dex (Z$=0.008$) 
or  ${\rm Fe/H]} = 0.0$\,dex (Z$=0.019$, i.e.  Solar). The systematics could be due to incorrect 
zero-point shifts applied to our estimates. Nevertheless, Fig.~\ref{fehyoung} shows that
we may have significantly underestimated the metallicities of young clusters. A close scrutiny 
of Fig.~\ref{fehyoung} also indicates that our estimates of [Fe/H] using the PEGASE-HR models 
are in slightly better agreement with those of Beasley et al. than those using the SSP models 
of Vazdekis et al. One possible explanation is that the SSP models of  Vazdekis et al. contain 
few templates of very young clusters. The lower age limit of templates of Vazdekis et al. is 
0.063\,Gyr, while many of the clusters in comparison here have ages less than 0.03\,Gyr
according to the determinations of Beasley et al. (2004).

Many more studies for old cluster are available in the literature and we compare our results with 
previous ones below. However, for completeness, old clusters analyzed by Beaslet et al. (2004) 
are  also show in Fig.~\ref{fehyoung}. The agreement with our values is excellent, 
except that our values derived using the PEGASE-HR models are slightly lower than 
those of Beasley et al.

\begin{figure*}
  \centering
  \includegraphics[width=0.98\textwidth]{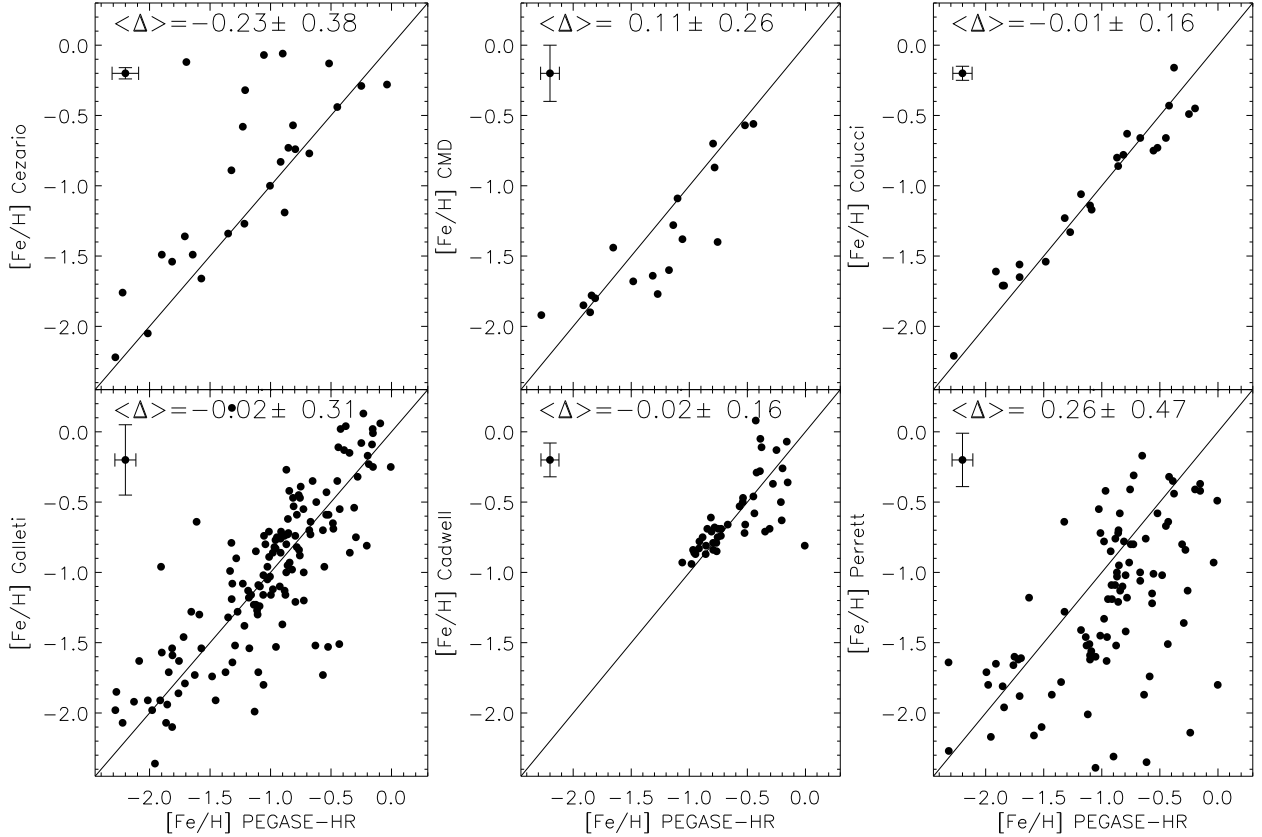}
  \caption{Comparisons of  our metallicity estimates obtained with the PEGASE-HR models 
  and those published in the literature for old clusters in M31. 
  The error bar in each panel represents the  median errors of measurements.}
  \label{fehsp}
  \label{fehph}
\end{figure*}

Now we compare our results for  old clusters in M31 with those in  the 
literature. For those old clusters  our results  derived with various  methods 
using  different SSP models are in good agreement. 
Because of this,  only values derived 
from the full spectral fitting using the  PEGASE-HR models are compared
to those in the literature. The comparisons are presented in Fig.~\ref{fehsp}.

We first  compare our [Fe/H] values with those obtained from 
low- and medium- resolution spectroscopy in the literature, 
i.e. those of \citet{Perrett2002, Galleti2009, Caldwell2011} and \citet{Cezario2013}. 
In general, the agreement is good.
\citet{Caldwell2011} estimated metallicities from spectra obtained with  
the 6.5\,m MMT Hectospec multi-fiber spectrograph. 45
objects of their targets  are in common with ours. 
The differences between our and their estimates have a small rms 
scatter of  0.16\,dex  and a negligible offset of $-$0.02\,dex.
The rms scatters of differences between our results  and those of 
\citet{Galleti2009} and \citet{Cezario2013} are slightly  larger.
Note, however, that the objects in common here cover a wider metallicity 
range than those in common with the sample of Caldwell et al. 
\footnote{{We have used the metallicities derived from 
the \eza~analysis in \citet{Caldwell2011}. Their metallicities, 
derived from a Lick index relation using measurements 
of metallicity [Fe/H] of Galactic GCs,, 
did cover the entire metallicity range, unlike the case for the \eza~values.}}, 
down to as low as $\sim -2$\,dex. 
\citet{Galleti2009} present metallicity 
estimates based on the Lick indices for 245 GCs in M31,
144 objects of them are in common with our sample. 
The rms scatter of the differences between the two sets of measurements is 
considerable, about  0.31\,dex.
The overall agreement is however very good, with no obvious systematics over 
a wide metallicity range. The average difference between the two sets of measurements 
amounts to only $-0.02$\,dex.
 \citet{Cezario2013} derive spectroscopic
metallicities of 38 M31 GCs by full spectral fitting. 
For the 27 objects in common with ours, the differences 
between their and our metallicity estimates have a mean of $-$0.23\,dex 
with a rms scatter of 0.38\,dex. 
\citet{Perrett2002} present spectroscopic metallicities of about 200 GCs in M31, 
derived from the Lick indices. For the 93 objects 
in common with ours, again covering a wide range of metallicities, 
the differences of their and our estimates have a mean of 0.26\,dex 
and a rms scatter of 0.47\,dex. Parts of the  large discrepancies 
between our results and those of \citet{Perrett2002} might be caused
by improper background subtraction in their work for clusters near  
the centre of  M31 where the background emission from the host galaxy  is high.
 
Probably the most robust  metallicity estimates  available hitherto for star clusters in 
M31 are those derived from the CMDs of the individual clusters based on  the mean 
colors of giant branch stars \citep{Rich2005,Fuentes2008,Perina2009,Caldwell2011}
and those determined from Fe~{\sc i} lines detected in 
high-resolution (R$\sim$20,000)  spectra \citep{Colucci2009, Colucci2014, Sakari2015}.
The giant branch stars of star clusters in M31 can be resolved by {\em HST} imaging  photometry.
We have collected metallicities derived from the CMDs of red giant branch stars in 
 the literature and found 18 clusters in common with 
 our sample (cf. middle panel of top row of Fig.~\ref{fehsp}). 
The objects come from \citet{Rich2005,Fuentes2008,Perina2009}
and \citet{Caldwell2011}. Metallicities based on high-resolution spectroscopy 
are collected from \citet{Colucci2009} and \citet{Colucci2014}. 
In total 24 objects in common  with our sample
are found  (cf. right panel of top row of Fig.~\ref{fehsp}). 
Fig.~\ref{fehsp}  shows surprisingly good agreements
 for both comparisons. Our measurements, compared to those derived from the 
 CMDs and from high-resolution spectroscopy, have an average difference of only 
0.11 and $-$0.01\,dex, respectively, along with a  rms scatter of 0.26 and 0.16\,dex, respectively.

\section{Ages}

\subsection{Ages from LAMOST spectra}

\begin{figure}
  \centering
  \includegraphics[width=0.48\textwidth]{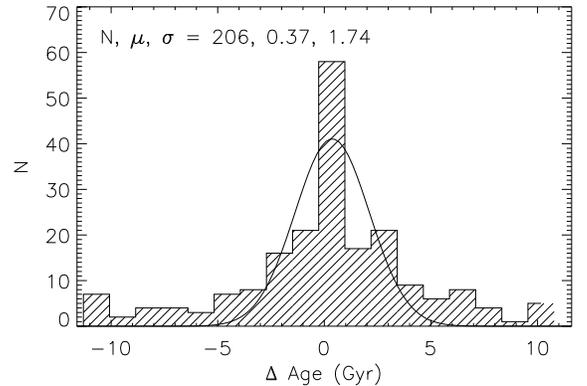}
  \caption{Same as Fig.~\ref{intfeh} but for  ages
  deduced from duplicate observations of the same targets.}
  \label{intage}
\end{figure}

\begin{figure}
  \centering
  \includegraphics[width=0.48\textwidth]{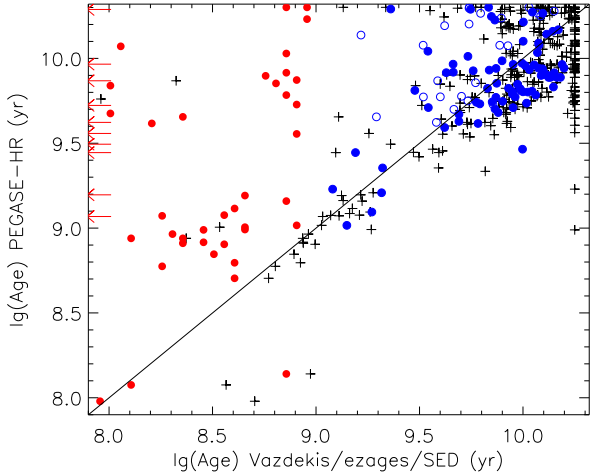}
  \caption{Comparison of cluster ages derived 
  with the PEGASE-HR models and those with the models of 
  Vazdekis et al. (black pluses), \eza~ (blue filled circles) and by SED 
  fitting (red filled circles). Blue open circles represent clusters  of 
  \eza~ ages but have  [Fe/H] $<-0.95$\,dex. Red arrows
mark clusters of SED ages smaller than 0.08\,Gyr.}  
  \label{sedezsp}
\end{figure}

In the process of  fitting the full spectrum using the ULySS code or  comparing the 
measured Lick line indices with  the predictions of SSP  models 
using the \eza~code, the best-fitted age of cluster is also determined 
simultaneously with the metallicity.
Ages of clusters thus derived with ULySS and \eza~are listed in Table~2.  

We first test the precision of ages delivered by full spectral fitting for clusters
with duplicate observations. Fig.~\ref{intage} shows that a precision of 
better than 2\,Gyr has been achieved for most clusters. 
Fig.~\ref{sedezsp} compares ages 
determined with the PEGASE-HR models and those with the models of  Vazdekis et al. In 
general, ages derived with the two sets of SSP models are consistent with each other. No 
obvious systematic offset is seen. The scatters of their 
differences are however considerable. We notice that some clusters 
of PEGASE-HR ages between $\sim$3 and $\sim$15\,Gyr 
are all found to have an age of  $\sim$ 15\,Gyr  as derived with models of 
the Vazdekis et al., i.e. the upper age limit of the models of Vazdekis et al. 
On the other hand, ages of those clusters returned by the \eza~code are consistent with 
the PEGASE-HR ages. It seems that  the full spectral fitting with the models of 
Vazdekis et al. may have overestimated the ages of some clusters, 
by returning an age of 15\,Gyr for those clusters,  the upper limit of age of the models. 

 The code \eza~only provide parameters for clusters 
 falling within  the model grids, as  Fig.~\ref{lickgrid} shows. 
This yields  ages of 103 old clusters in our sample. The results are 
 compared with those yielded by  ULySS fitting with the PEGASE-HR models in 
 Fig.~\ref{sedezsp}. The agreement is good overall, with a negligible average 
offset of $\sim$1\,Gyr. Note that  ages yielded by  \eza~are based on the 
 H$\beta$ index only.  
{The Padova isochrones, adopted in the models of \citet{Schiavon2007} and 
used in \citet{Caldwell2011}, do not contain blue horizontal branch (HB) stars. 
Because an actual metal-poor cluster with a blue HB tends to have stronger 
Balmer lines than predicted by Padova models of the same age and 
metallicity, cluster ages yielded by EZ Ages tend to be underestimated 
for metal poor clusters.}
 This effect can be clearly seen in Fig.~\ref{sedezsp}, 
 where \eza~ ages of clusters of [Fe/H]$ 􏰍<-0.95$\,dex are all smaller than those derived with  
ULySS with the PEGASE-HR models. 
If  we exclude those  clusters of [Fe/H]$ 􏰍<-0.95$\,dex, 
then no obvious systematic offset is seen between the \eza~ and PEGASE-HR ages.

\subsection{Ages from SED fitting}

\begin{figure}
  \centering
  \includegraphics[width=0.48\textwidth]{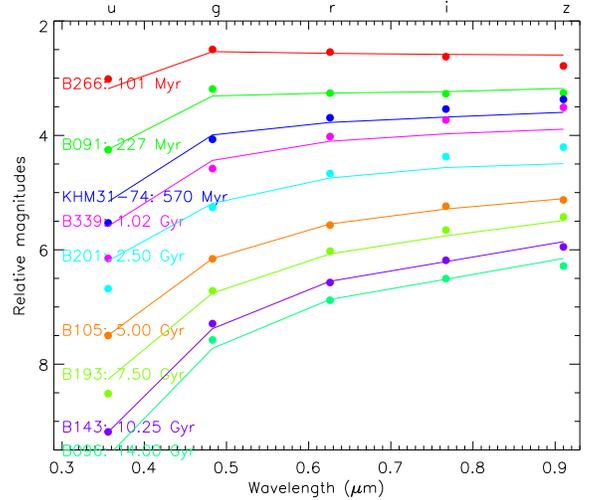}
  \caption{Examples of SED fitting for $ugriz$ photometry. 
  Filled circles are photometric 
  measurements in different passbands and solid lines are from the SSP
  models. Object names and estimated ages marked. 
  Different colors represent different objects. 
The magnitudes of a given different object  have been shifted for clarity.}
  \label{sedfit}
\end{figure}

\begin{figure*}
  \centering
  \includegraphics[width=0.78\textwidth]{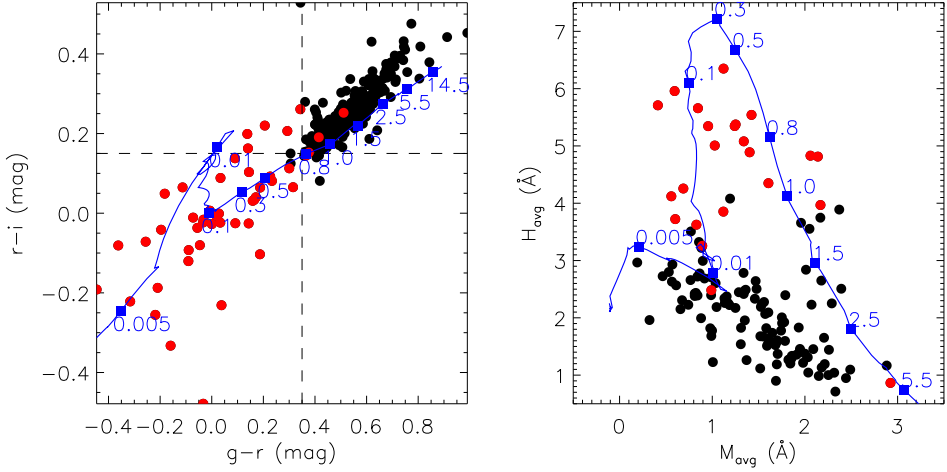}
  \caption{$g-r$ versus $r-i$ (left panel) color-color  
  and $M_{avg}$ versus $H_{avg}$ (right panel) line index-index diagrams 
  of M31 star clusters. Red and black  filled circles represent  
  young ($<$1\,Gyr) and old ($> 1$\,Gyr) clusters, respectively. 
  The solid curves  delineates the locus of clusters of different ages 
  between 0.001 and 14.5\,Gyr from the SSP models of \citet{Bruzual2003} SSP models.
The curve is marked by ages in Gyr.
The dashed lines in the left panel divide young and old clusters in colors.}
  \label{ageccd}
\end{figure*}

Ages of young clusters can be estimated by SED fitting of multi-band 
photometry \cite[and references therein]{Wang2010, Fan2010, Kang2012}. 
We have  collected magnitudes in $ugriz$ passbands for clusters 
in our sample (see Sect.~2.3). With photometric data 
from the optical to the NIR, we are able to break the age-metallicity degeneracy 
of young clusters \citep{Anders2004}. The photometric data 
are de-reddened using  reddening values described in \S{2.3} and  
the reddening law  of  \citet{Yuan2013}. 
The de-reddened data are then fitted with predictions of the  
SSP models to determine the ages. The SSP models of 
\citet[hereafer BC03]{Bruzual2003}\footnote{http://www2.iap.fr/users/charlot/bc2003/} 
are adopted. BC03 models are constructed using various 
stellar libraries and IMFs. In this work we adopt those models 
of IMF from \citet{Chabrier2001} and stellar library from the Padova isochrones. 
The models cover ages of $5.0< {\rm log}\,t < 10.3$\,(yr), 
with log\,$t$ bins ranging from 0.005 to 0.05\,(yr). The models have six metallicities
(of values of Z of 0.0001, 0.0004, 0.004, 0.008, 0.02 and 0.05). 

The SED fitting is applied to all sample star clusters 
that are detected in at least four bands. 
For each object, we select BC03 model of metallicity closest to  that
derived for that object with full spectral fitting. The age is then determined by  
 minimising  $\chi^2$ defined as,
\begin{equation}
  \chi^2=\sum\frac{(m^{\rm obs}_{\lambda i}-m^{\rm mod}_{\lambda i}(t))^2}{\sigma_{\lambda 
  i}^2},
\end{equation}
where $m^{\rm obs}_{\lambda i}$ and $m^{\rm mod}_{\lambda i}$ are the 
observed and model magnitudes in band $\lambda i =u, ~g, ~r, ~i$ or $z$;
and $\sigma_{\lambda i}$ is the uncertainty of magnitudes 
in band $\lambda i$. The uncertainty include the contributions from
observed and model magnitudes as well as from the error of distance modulus \citep{Ma2012b}, i.e., 
$\sigma^2_{\lambda i}=\sigma^2_{{\rm obs},\lambda i}+\sigma^2_{{\rm mod},\lambda i}+\sigma^2_{{\rm dm},\lambda i}$. 
Note that  $\sigma _{{\rm mod},\lambda i}$ and $\sigma _{{\rm dm},\lambda i}$ 
only affect the absolute value of $\chi^2$ but not the best-fit result. In the current work,
we have assumed $\sigma_{{\rm mod},\lambda i} =0.05$\,mag and $\sigma_{{\rm dm},\lambda i} 
=0.07$\,mag.

Ages of 292 clusters in our sample are obtained with the 
SED fitting. They are listed in Table~3. 
Examples of best-fits for nine objects  are plotted in Fig.\ref{sedfit}. 
Fig.\ref{sedfit} shows that the SEDs of old clusters have 
similar shapes that are hard to distinguish. Thus 
the SED best-fit ages are only adopted for young clusters.
In total there are {46 young clusters} with best-fit SED ages $t \leq $1\,Gyr in our sample. 
The left panel of Fig.~\ref{ageccd} plots the dereddened $g-r$ versus  $r-i$
 color-color diagram of all clusters in our sample. 
Young and old clusters are well separated in 
 color $g-r$, with young ones having $g-r < 0.35$\,mag. 
 This is consistent with the color criteria ($g-r < 0.3$\,mag) that
 \citet{Peacock2010} used to select young clusters.
 The cut  between young and old clusters in color $r-i$ is less clear. 
A roughly  border line is $r-i = 0.15$\,mag, although there are 
young and old clusters on both side of the line.
In the right panel of Fig.~\ref{ageccd} 
we present a plot of the average Balmer line index 
H$_{\rm avg}$  versus the average  metal line index  M$_{\rm avg}$ measured
from the LAMOST spectra of clusters with a \bsnr$>$10 in our sample.
The indexes are defined as $\rm M_{avg} = (Fe5270+Mgb)/2$ and 
$\rm H_{avg} = (H\delta F + H\gamma F + H\beta)/3$, respectively. 
Except for B016, all young clusters identified by SED fitting fall  around 
the sequence line for young clusters. 
For B016 the  SED fitting yields  log\,$t=8.9\pm 0.2$\,(yr), significantly younger than 
log\,$t=10\pm0.2$\,(yr) given by  full spectral 
fitting. The latter  is consistent with the estimate of  \citet{Caldwell2011}, log\,$t=10$\,(yr). 
The age of B016 from SED fitting might have been underestimated but the value is still consistent with
that from the full spectral fitting considering the uncertainties of both estimates.

 \begin{figure}
  \centering
  \includegraphics[width=0.49\textwidth]{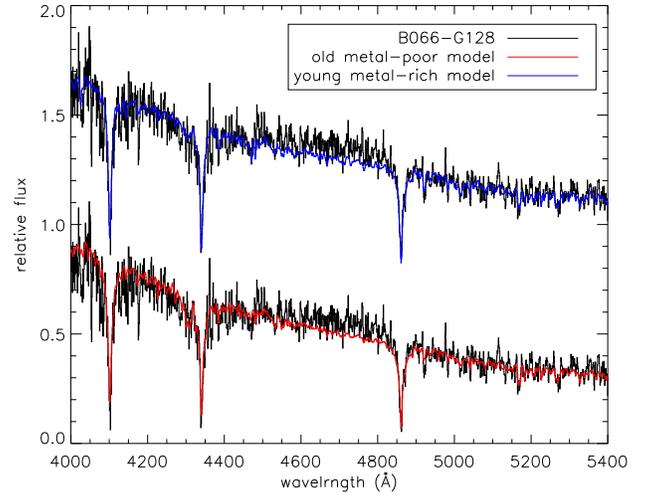}
    \caption{{The LAMOST spectra of a young and metal-rich cluster, B066-G128, compared 
    with two SSP model spectra from \citet{Vazdekis2010}. 
   The spectrum of B066-G128 is plotted in black and  
   the spectra of the Vazdekis et al. models are plotted in red (with model parameters 
    [Fe/H] = $-$1.71\,dex and $t$=1.122\,Gyr) and blue (with model parameters 
    [Fe/H] = 0.00\,dex and $t$ = 0.0708\,Gyr), respectively. The continuum spectra of
     those two spectra from the Vazdekis et al. models are  
     normalised to that of B066-G128.  } }
  \label{ymcspec}
\end{figure}

Ages derived from the various approaches have already been compared 
in Fig.~\ref{sedezsp}, including those from the SED fitting for young clusters.
The scatters are significant. Ages derived from full spectral fitting are systematically 
older than those from SED fitting. Some of the young 
clusters identified by SED fitting have very old ages, $3-15$\,Gyr, from full spectral fitting.
Such old ages are inconsistent with the colors and  
Balmer/metal line indices of those young clusters. We believe  that full spectral
fitting may have overestimated the ages of young clusters. 

{
We have checked carefully the spectra of those young clusters as well as  
those of both the PEGASE-HR and Vazdekis et al. models. 
The young clusters are found to have strong Balmer lines and
weak metal lines (such as Mgb line). They can be fitted by not only
young and metal-rich SSP models, but also old and metal-poor
models. An example is shown in Fig.~\ref{ymcspec}. B066 is a young cluster, with
$t$ = 0.72\,Gyr from the SED fitting method in the current work and 
$t$ = 0.71\,Gyr from \citet{Perina2010}. The metallicity of B066 is 0.0\,dex from \citet{Perina2010}.
Yet the full spectral fitting method gives values of [Fe/H] = $-$2.12 and   $-$1.52\,dex and
$t$ = 1.17 and 1.07\,Gyr from PEGASE-HR and Vazdekis et al. models, respectively. 
Both the old, metal-poor model 
and the young, metal-rich model 
fit well with the observed spectra of B066.
Thus there is degeneracy when we use the full spectral fitting method. 
In most cases, young clusters have been assigned old ages along with poor metallicities, 
such as VDB0, B448, B342 and B066, etc. Both ages and metallicities of these
clusters obtained from the full spectral fitting are problematic. 
For these clusters we adopt only the ages from SED fitting
and simply assigned their metallicities as 0.0\,dex.
We have also found two opposite cases 
(B074 and B013). They are old, metal-poor clusters, but have been assigned a young 
age and a high metallicity based on the full spectral fitting method. 
B074 is misclassified by the Vazdekis et al. models, but correctly classified by the 
PEGASE-HR models. B013 is misclassified by the PEGASE-HR models but 
correctly classified by the Vazdekis et al. models.
The abnormal results of these two clusters are excluded in Table~3.}
 
\subsection{Comparison with previous age estimates}

 \begin{figure*}
  \centering
  \includegraphics[width=0.99\textwidth]{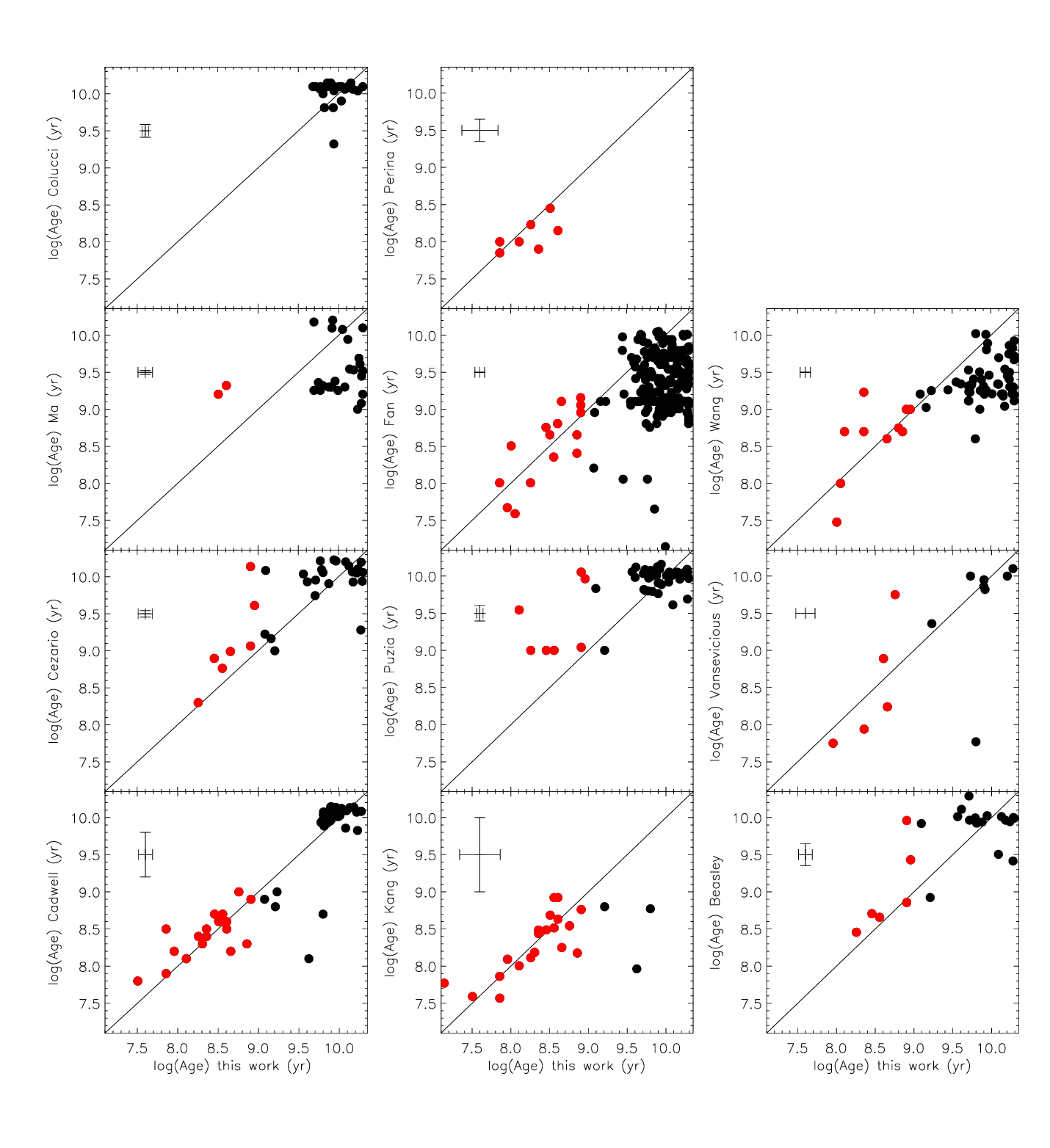}
  \caption{Comparisons of ages from our analysis and those in the literature:
  \citet{Beasley2004},
  \citet{Perina2010},   \citet{Puzia2005},   \citet{Vansevicius2009}, 
  \citet{Caldwell2009, Caldwell2011},   \citet{Colucci2009},   \citet{Colucci2014}, 
  \citet{Ma2009, Ma2011},   \citet{Ma2012},   \citet{Wang2010},   \citet{Fan2010}, 
   \citet{Kang2012} and \citet{Cezario2013}. Our ages derived from full spectral fitting using 
 PEGASE-HR SSP models for old clusters are plotted in black 
 and those from SED fitting for young clusters 
 are plotted in red. The error bars in each panel show the median errors of age determinations.}
  \label{agecmp1}
  \label{ageph}
\end{figure*}

Ages of old clusters derived from the full spectral fitting with the PEGASE-HR models 
are in good agreement with those from \eza~, while the PEGASE-HR ages for young 
clusters are probably overestimated. For the purpose of comparisons with 
previous work, we have adopted the PEGASE-HR ages for old clusters in our sample,
while for young clusters the ages from SED
fitting are used when available. Note that due to lack of suitable photometric data, 
 SED fitting does not
provide ages for all young clusters in the  sample. 
The number of young clusters without a SED fitting age is however small 
($\leq$~5). For those a couple of clusters, their ages yielded by  full spectral fitting 
may have been  overestimated (see Fig.~\ref{agecmp1}).

In Fig.~\ref{agecmp1}, we compare our ages with those in 
the literature,  \citet{Beasley2004},
  \citet{Perina2010},   \citet{Puzia2005},   \citet{Vansevicius2009}, 
  \citet{Caldwell2009, Caldwell2011},   \citet{Colucci2009},   \citet{Colucci2014}, 
  \citet{Ma2009, Ma2011},   \citet{Ma2012},   \citet{Wang2010},   \citet{Fan2010}, 
   \citet{Kang2012} and \citet{Cezario2013}.
\citet{Beasley2004} derive ages of 30 clusters in M31 from spectral indices. 
24 of them are in common with our sample. Their results are very similar with 
ours. The scatter is small. 
\citet{Puzia2005} present ages of 70 GCs of M31 from the Lick line indices. 
50 of them are found in our sample. In general the agreement between our 
age  estimates  and theirs is good, except for 6 young clusters. Considering the SSP 
models used by \citet{Puzia2005} do not cover ages younger than $1$\,Gyr, their 
 estimates for young clusters may be incorrect.
\citet{Vansevicius2009} derive ages of 238 high probability star cluster candidates in M31
selected based on multi-band photometric data and images. 17 of them are included in 
our sample. Our results are consistent with theirs, except for one GC, 
B384, for which we derive an age of log\,$t=9.8$\,(yr) while \citet{Vansevicius2009}  
give a result of log\,$t$=7.77\,(yr). Our result is consistent with 
that of \citet{Caldwell2011}  who find  log$\,t=10.1$\,(yr). 
\citet{Caldwell2009} estimate ages of young clusters from spectral indices and
\citet{Caldwell2011} derive ages of old clusters from spectra indices  
using \eza. 73 objects are common between our 
sample and those of  \citet{Caldwell2009,Caldwell2011}. The 
agreement between our and their  estimates  is very good.
The ages of old clusters from \citet{Caldwell2011} are slightly  larger 
than our results. This could be caused by the different SSP models used to  
calculate the ages. Note that  \citet{Caldwell2011} simply assign an age of 14\,Gyr to 
all clusters of  [Fe/H] $< -$0.95\,dex. This may not only lead to overestimated ages,
but also masses of some clusters (see \S{5}). 
Our ages of five clusters (B124, B338, B119, B106 and B365)
found to be young by \citet{Caldwell2009}
are overestimated. The differences for three of them
(B124, B119 and B106) are however quite small, in fact  consistent within the error bars.
The differences for the other two objects (B338 and B365) are slightly larger.
Both clusters are also found to be young by \citet{Kang2012}, who 
estimate ages of 182 young clusters by SED fitting of  multi-band photometric data.
A comparison of  our age estimates  and  those of  
\citet{Kang2012} is also presented in Fig.~\ref{agecmp1}.
The agreement  is good  except for  B338 and B365.
\citet{Cezario2013} derive ages by  full spectral fitting using 
code ULySS and SSP models of Vazdekis et al. 
For old clusters it is not surprisingly that their results are in good agreement with ours. 
For  young clusters, the estimates of 
 \citet{Cezario2013}, as those of ours based on full spectral fitting, 
  are systematically higher than our values derived from SED fitting and adopted here.
 \citet{Perina2010} estimate ages of some young clusters by comparing 
model isochrones in  CMDs with data obtained from the {\em HST}/WFPC2 observations.
7 objects are in common with our sample and 
the agreement is again good. 
\citet{Colucci2009} and \citet{Colucci2014} determine ages for some star 
clusters with  high-resolution spectroscopy. 24 objects studied by them are included in our 
sample. The comparison shows that our estimates  
are consistent with results from high-resolution spectroscopy.
\citet{Ma2009, Ma2011, Wang2010} and \citet{Fan2010} estimate ages of M31 
star clusters by SED fitting of  multi-band photometric measurements. For young clusters, their results are 
consistent with ours. However,  their estimates  of old clusters 
are systematically smaller than ours. The discrepancies are significant. 
As discussed above, the SED fitting may  not be suitable for  determining ages of old clusters. 

\section{Masses}

\begin{figure}
  \centering
  \includegraphics[width=0.48\textwidth]{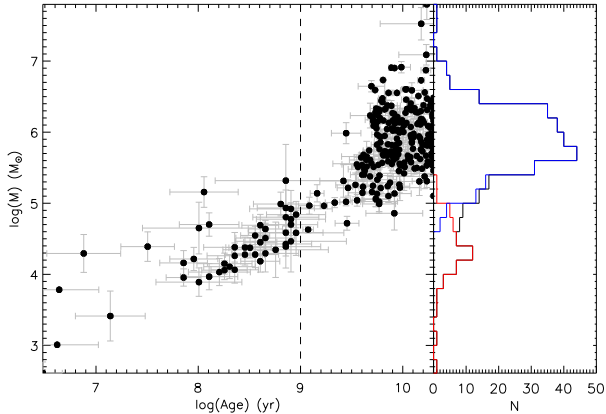}
  \caption{Masses estimated for  M31 star clusters in our sample
plotted against ages of the clusters. 
 The vertical dashed line at age  1\,Gyr  separates the young and old clusters.
A histogram of masses of all clusters is also plotted on the right. Black, red and
green lines give respectively the  mass distributions of all,  
young and old clusters.}
  \label{agemass}
\end{figure}

\begin{figure*}
  \centering
  \includegraphics[width=0.78\textwidth]{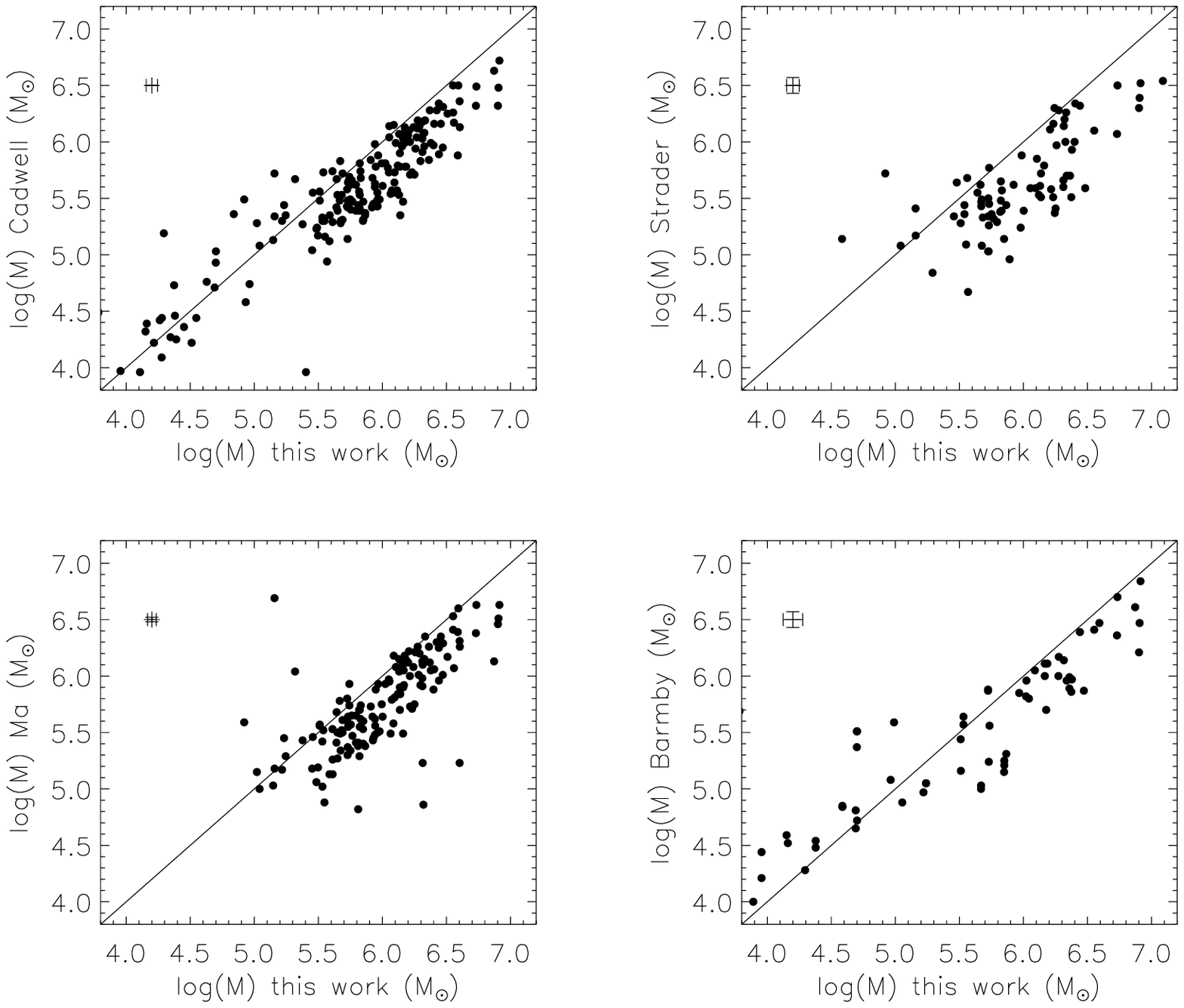}
  \caption{Masses obtained in the current work are compared with those published by 
 \citet{Barmby2007, Barmby2009},  
 \citet{Caldwell2009, Caldwell2011},  \citet{Strader2011}, 
  and by \citet{Ma2015}. The error bars in each panel show the  
  median errors of mass determinations.}
  \label{masscp}
\end{figure*}

Once the age and metallicity of a star cluster  have been determined, 
its mass  can be estimated by   comparing 
the photometric measurements  with the SSP models. In the current work, we 
use  the SDSS optical $ugriz$ photometry and the BC03 models for the purpose. The 
BC03 models are normalized to one Solar mass  in stars of age $t$=0\,Gyr. 
Thus the mass of a cluster can be 
estimated by the difference between the observed intrinsic absolute magnitude and 
that predicted  by the models in $ugriz$ bands. To calculate the intrinsic absolute 
magnitudes of clusters in our sample, we use the reddening values described in 
\S{2.3}  and listed in Table~1. The distance modulus of M31 is taken to be 
$(m - M)_0$ = 24.43\,mag \citep{Caldwell2011}.
Clusters metallicities derived from full spectral fitting 
with the PEGASE-HR models are adopted  when 
calculating the masses. 
As for the ages, again those yielded by the PEGASE-HR models are adopted except
for young clusters for which the ages from SED fitting are used when available.

With this approach, we have been able to  
determine masses of  295 clusters in our sample. The results are  listed in Table~3. 
In Column `Note' of Table~3, the method used to derive the ages adopted  
for calculating the mass is marked, where values 1 and 2 refer to 
PEGASE-HR and SED fitting ages, respectively.
The  masses thus estimated  for our
M31 sample  clusters are plotted against age in Fig.~\ref{agemass}. 
The estimated masses  span a range from $\sim 10^3$ to $\sim 10^7{\rm 
M_\odot}$. Two discrete peaks are seen in the cluster mass histogram distribution,
at $10^{4.3}$ and $10^{5.7}\,{\rm M_\odot}$, 
produced by young and old clusters in the sample, respectively. 

We compare in Fig.~\ref{masscp} our mass  estimates to 
determinations in the literature 
(\citealt{Barmby2007, Barmby2009, Caldwell2009, Caldwell2011, Strader2011, Ma2015}). 
The masses presented in \citet{Barmby2007, Barmby2009}, \citet{Caldwell2009,Caldwell2011} 
and \citet{Ma2015} are all estimated by adopting certain values of mass-to-light ratio 
$M/L$. \citet{Barmby2007, Barmby2009} 
and \citet{Ma2015} use $M/L$ ratios given by the SSP 
models as in the case of the current work. 
\citet{Caldwell2011}  assume constant ratio of  
$M/L_V= 2$. In contrast, the masses presented in 
\citet{Strader2011} are estimated from the velocity dispersions 
and structural parameters of the star clusters. 
Overall, our mass estimates are in good agreement with 
determinations presented in those studies. For old ($t>1$\,Gyr) or 
massive ($M >10^{5}\,{\rm M_{\odot}}$) clusters
our mass estimates are slightly larger than the previous determinations. This is probably  
caused by the different SSP models used. 
The fact  that both \citet{Caldwell2011} and \citet{Ma2015}
calculated the cluster masses using  the ages of \citet{Caldwell2011}, which 
are  slightly higher than our values  (see \S{4.3}), 
may also partly be responsible for the discrepancies.

\begin{table*}\addtolength{\tabcolsep}{-2pt}
 \centering
  \caption{Derived properties of star clusters in our sample.}
  \begin{tabular}{lrrrrrrrrrrr}
  \hline
  \hline
Name &   [Fe/H]$^a$ & [Fe/H]$^b$ &  [Fe/H]$^c$ &  [Fe/H]$^d$ &
[Fe/H]$^e$ & log\,$t$$^f$ & log\,$t$$^g$ & log\,$t$$^h$ & log\,$t$$^i$ & Note$^j$ & 
log\,$M$  \\
  & (dex) & (dex) & (dex) & (dex) & (dex) & (yr) & (yr) & (yr) & (yr) & & ($\rm M_{\odot}$)  \\
 \hline
 B001-G039 &   $  -0.66 \pm    0.11  $ &   $  -0.84 \pm    0.04  $ &   $  -0.80  $ &   $  -0.75  $ &   $  -0.88  $ &   $   9.80 \pm    0.15  $ &   $  10.25 \pm    0.01  $ &   -- &   $   9.95  $ &  1 &   $   5.69 \pm    0.12  $ \\
 B002-G043 &   $  -2.17 \pm    0.24  $ &   $  -2.32 \pm    0.01  $ &   -- &   -- &   -- &   $   9.86 \pm    0.11  $ &   $  10.05 \pm    0.19  $ &   -- &   -- &  1 &   $   5.10 \pm    0.09  $ \\
 B003-G045 &   $  -1.70 \pm    0.13  $ &   $  -1.33 \pm    0.15  $ &   $  -1.42  $ &   -- &   -- &   $  10.24 \pm    0.05  $ &   $   9.62 \pm    0.08  $ &   -- &   -- &  1 &   $   5.59 \pm    0.04  $ \\
 B004-G050 &   $  -0.65 \pm    0.06  $ &   $  -0.73 \pm    0.06  $ &   $  -0.58  $ &   $  -0.71  $ &   $  -0.75  $ &   $  10.27 \pm    0.02  $ &   $  10.19 \pm    0.07  $ &   -- &   $  10.10  $ &  1 &   $   5.93 \pm    0.02  $ \\
 B005-G052 &   $  -0.63 \pm    0.04  $ &   $  -0.78 \pm    0.02  $ &   $  -0.69  $ &   $  -0.63  $ &   $  -0.87  $ &   $   9.92 \pm    0.06  $ &   $  10.25 \pm    0.01  $ &   -- &   $  10.15  $ &  1 &   $   6.26 \pm    0.05  $ \\
 B006-G058 &   $  -0.49 \pm    0.04  $ &   $  -0.52 \pm    0.04  $ &   $  -0.28  $ &   $  -0.41  $ &   $  -0.28  $ &   $   9.92 \pm    0.06  $ &   $  10.00 \pm    0.06  $ &   -- &   $   9.66  $ &  1 &   $   6.19 \pm    0.05  $ \\
 B008-G060 &   $  -0.59 \pm    0.09  $ &   $  -0.76 \pm    0.06  $ &   $  -0.70  $ &   $  -0.73  $ &   $  -0.63  $ &   $   9.92 \pm    0.13  $ &   $  10.25 \pm    0.05  $ &   -- &   $   9.63  $ &  1 &   $   5.73 \pm    0.11  $ \\
 B010-G062 &   $  -1.53 \pm    0.15  $ &   $  -1.38 \pm    0.15  $ &   -- &   -- &   -- &   $   9.69 \pm    0.10  $ &   $   9.69 \pm    0.09  $ &   -- &   -- &  1 &   $   5.56 \pm    0.07  $ \\
 B011-G063 &   $  -1.49 \pm    0.09  $ &   $  -1.37 \pm    0.15  $ &   $  -1.00  $ &   $  -1.37  $ &   $  -0.95  $ &   $  10.19 \pm    0.03  $ &   $   9.93 \pm    0.12  $ &   -- &   $   9.62  $ &  1 &   $   5.82 \pm    0.02  $ \\
     B011D &   $  -1.17 \pm    0.09  $ &   $  -1.18 \pm    0.12  $ &   $  -0.91  $ &   $  -1.36  $ &   $  -0.81  $ &   $   9.62 \pm    0.08  $ &   $   9.73 \pm    0.12  $ &   -- &   $   9.58  $ &  1 &   $   5.40 \pm    0.06  $ \\
 B012-G064 &   $  -2.02 \pm    0.03  $ &   $  -1.91 \pm    0.02  $ &   $  -0.16  $ &   $  -1.61  $ &   $  -0.12  $ &   $  10.08 \pm    0.04  $ &   $  10.07 \pm    0.04  $ &   -- &   $   9.52  $ &  1 &   $   6.40 \pm    0.04  $ \\
 B013-G065 &   -- &   $  -0.59 \pm    0.10  $ &   -- &   -- &   -- &   -- &   $  10.05 \pm    0.17  $ &   --  &   -- &  1 &   $   5.67 \pm    0.15  $ \\
... & ... & ... & ... & ...& ... & ... & ... & ... & ... & ... & ... \\
  KHM31-74 &   $  -0.62 \pm    0.12  $ &   -- &   -- &   -- &   -- &   -- &   -- &   $   8.76 \pm    0.52  $ &   -- &  2 &   $   4.35 \pm    0.39  $ \\
  LAMOST-1 &   $  -0.31 \pm    0.04  $ &   $  -0.36 \pm    0.03  $ &   $  -0.29  $ &   $  -0.06  $ &   $  -0.29  $ &   $   9.97 \pm    0.05  $ &   $  10.16 \pm    0.04  $ &   -- &   $   9.66  $ &  1 &   $   5.26 \pm    0.04  $ \\
  LAMOST-2 &   $  -0.57 \pm    0.08  $ &   $  -0.66 \pm    0.08  $ &   -- &   -- &   -- &   $  10.16 \pm    0.31  $ &   $  10.25 \pm    0.01  $ &   -- &   -- &  1 &   $   5.31 \pm    0.20  $ \\
  LAMOST-3 &   $  -1.58 \pm    0.12  $ &   $  -1.75 \pm    0.13  $ &   $  -1.54  $ &   $  -1.60  $ &   -- &   $   9.78 \pm    0.07  $ &   $  10.07 \pm    0.16  $ &   -- &   -- &  1 &   $   5.22 \pm    0.06  $ \\
  LAMOST-4 &   $  -1.90 \pm    0.20  $ &   $  -1.86 \pm    0.13  $ &   $  -1.34  $ &   $  -1.45  $ &   -- &   $  10.30 \pm    0.01  $ &   $  10.09 \pm    0.16  $ &   -- &   -- &  1 &   $   5.61 \pm    0.01  $ \\
  LAMOST-5 &   $  -2.18 \pm    0.04  $ &   $  -2.17 \pm    0.03  $ &   $  -2.36  $ &   $  -2.00  $ &   -- &   $  10.17 \pm    0.03  $ &   $   9.95 \pm    0.05  $ &   -- &   -- &  1 &   $   6.11 \pm    0.02  $ \\
      M086 &   $  -0.28 \pm    0.16  $ &   -- &   -- &   -- &   -- &   -- &   -- &   $   8.61 \pm    0.19  $ &   -- &  2 &   $   4.18 \pm    0.15  $ \\
      ... & ... & ... & ... & ...& ... & ... & ... & ... & ... & ... & ... \\
              \hline
\end{tabular}
  \label{ta3}
\begin{flushleft}
 This is a sample of the full table, which is available in its entirety in the electronic versions of this article.\\ 
$^a$ Clusters metallicities from full spectral fitting with the  PEGASE-HR models. \\
$^b$ Clusters metallicities from full spectral fitting with the models of  Vazdekis et al. \\
$^c$ Clusters metallicities from the [MgFe] index using the relation of  \citet{Galleti2009}. \\
$^d$ Clusters metallicities from the \avfe~index using the relation of  \citet{Caldwell2011}. \\
$^e$ Clusters metallicities from the \eza~package. \\
$^f$ Cluster ages from full spectral fitting with the  PEGASE-HR models. \\
$^g$ Cluster ages from full spectral fitting with the models of  Vazdekis et al. \\
$^h$ Cluster ages from SED fitting. \\
$^i$ Cluster ages from the \eza~package. \\
$^j$ Note$=$1: Mass estimated using the PEGASE-HR age; 
2: Mass estimated using age from  SED fitting.
\end{flushleft}
\end{table*}

\section{Discussion}

\subsection{GCs in streams}

The stellar halo of M31 is rich of substructures including streams, 
loops and filaments. Both metal rich and poor substructures are 
detected (e.g. \citealt{Ibata2014}), pointing to the presence of ongoing  multiple accretion events.
Many GCs found at large distances from the centre of M31 appear spatially associated with those 
substructures \citep{Mackey2010}. Some 
of those distant clusters are included in the current  sample. 
The most intriguing case  is LAMOST-1,
a new identified GC with  LAMOST (Paper I). It falls on
the Giant Stellar Stream (${\rm [Fe/H]} > -0.6$\,dex; \citealt{Ibata2014}) 
and has a  radial velocity suggesting its association with 
the Stream. In the current  work, we find a value of 
metallicity [Fe/H]$=-0.4$\,dex and an age of 9.2\,Gyr for LAMOST-1. 
LAMOST-1 is the most metal-rich of amongst  the distant  clusters in our sample. 
Its derived metallicity is compatible to that of  RGB stars in the Giant Stellar Stream 
\citep{Tanaka2010,Ibata2014}. The deduced age of LAMOST-1 is also in good agreement with 
the average age of red clump (RC) stars in the Giant Stellar Stream \citep{Brown2006,Tanaka2010}. 
The metallicity and age derived of 
LAMOST-1 suggest that it probably 
formed in an early-type and relatively massive dwarf galaxy, 
of mass comparable to that of  the Large Magellanic Cloud (LMC) or 
the Sagittarius (Sgr) dwarf spheroidal.

The other remote GCs at large distances from M31 
in our sample are all very old, comparable to the age of the universe, suggesting parent 
galaxies formed in the early universe. Those clusters have however a range of  metallicities,
suggesting the variety of substructures currently being accreted into the outer halo of M31.
MCGC8 is the second most metal-rich distant  GC in our sample. It falls on
Stream D in the halo of  M31. The metallicity of MCGC8 is [Fe/H]$=-$1.1\,dex, 
similar to values found for  stars in the stream \citep{Tanaka2010}.
It may be formed in a fairly massive galaxy, considering its metallicity. 
H26 falls on Stream C and has a metallicity [Fe/H]$=-1.6$\,dex, again  
consistent with values determined for stars associated with the 
stream \citep{Tanaka2010}. 
The relatively low metallicity suggests that the stream may be the relics of 
an intermediate-mass dwarf galaxy. 
LAMOST-4 and G002 fall on `Association 2',  a spatial overdensity of GCs found 
by \citet{Mackey2010}. They have very similar metallicities ([Fe/H] = $-$1.9\,dex for
LAMOST-4 and $-$2.1\,dex for G002) and are spatially close to each other.
Thus they may both come from a  dwarf galaxy of a relatively low mass.

\subsection{Age and metallicity distributions}

\begin{figure*}
  \centering
  \includegraphics[width=0.78\textwidth]{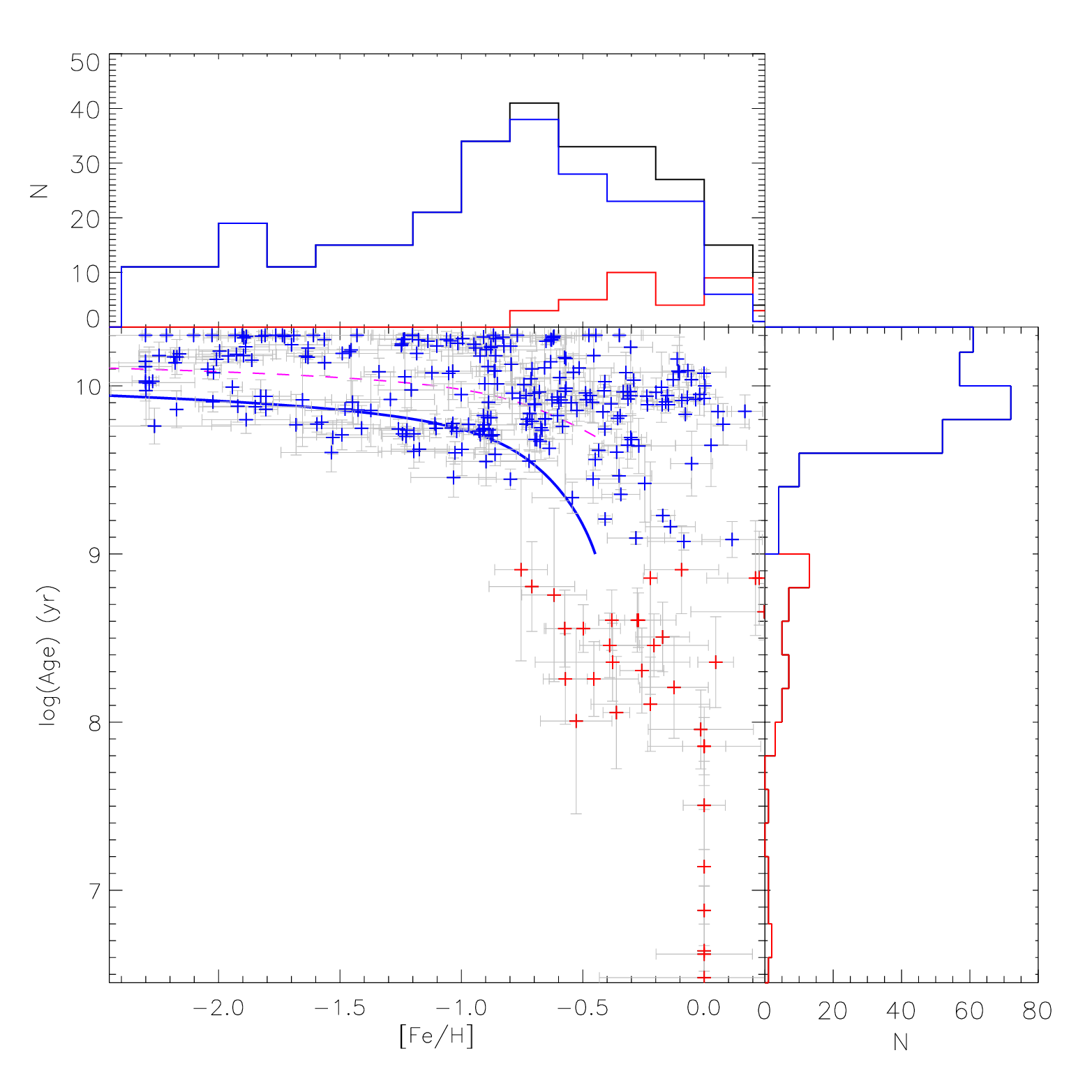}
  \caption{Ages plotted against metallicity for M31 clusters in our sample. 
Histograms of metallicity and age distributions are also plotted on the sides. 
Red and blue pluses (with grey error bars) represent young 
and old clusters in our sample, respectively. 
The black, red and blue histograms give distributions of all, young and old clusters, respectively. 
The pink dashed line delineates  age-metallicity relation 
of (accreted) Galactic GCs associated with the 
Sgr and CMa dwarf galaxies from \citet{Forbes2010}. The blue line is the 
same relation but shifted by 4\,Gyr.}
  \label{fehage}
\end{figure*}

The ages derived for  all clusters in our sample 
are plotted against metallicities in Fig.~\ref{fehage}. 
Here the metallicities are those obtained from full spectral fitting with the
PEGASE-HR models. For old clusters, the values  are consistent with estimates derived 
from other methods as well as with previous determinations in the 
literature. For young clusters, they may have been  
underestimated, by less than 0.5\,dex compared to 
those of \citet{Beasley2004} and \citet{Perina2010}. 
Ages for most objects in the plot are again those derived with the PEGASE-HR 
models, except for young clusters for which the  ages deduced by 
SED fitting are adopted when available. 
A small number  of the young clusters have very low metallicities with [Fe/H] $<-1.0$\,dex.
The values may be underestimated given that  the PEGASE-HR models 
are not suitable for  young clusters, as 
discussed in \S{4.2}. If we ignore those objects, then 
the remaining  young clusters in our sample clump at the 
bottom right of the Figure, with similar ages around 0.3\,Gyr and metallicities around 
$-0.3$\,dex. The latter is comparable to the solar value considering that we may have  
underestimated the metallicities of those young clusters. 
The result is consistent with previously studies (e.g. \citealt{Beasley2004}).

Most old clusters in our sample, 
240 in number, have ages larger than 
4\,Gyr. Only 19 have ages in the range of $1< t < 4$\,Gyr. The 
ages of old clusters peak  at $t \sim$8\,Gyr.
The metallicities of Galactic GCs are known to have 
a bimodal Gaussian distribution, peaking at  [Fe/H]=$-$1.60 and 
$-$0.59\,dex, respectively \citep{Harris1996, Galleti2009}.
The distribution of old clusters of M31 in our sample is quite different.
It has a very broad than bimodal  distribution. Again this 
is consistent with the findings of most 
previous studies \citep[and references therein]{Barmby2000,Perrett2002,
Lee2008,Galleti2009,Caldwell2011}. In any case,  the metallicity distribution of 
M31 GCs resembles nothing like  a single Gaussian distribution. 
It has an obvious broad peak around [Fe/H]=$-0.7$\,dex,  
 but also an almost flat distribution down to $\sim -2$\,dex.

The old clusters in Fig.~\ref{fehage} can be loosely divided into three groups.
The group of metallicities richer than $\sim -0.7$\,dex
have a wide range of ages (1--15\,Gyr). 
Objects poorer than [Fe/H] $\sim -0.7$\,dex seem to have two 
groups  -- one of the oldest ages with all values of 
metallicity down to $\sim -2$\,dex and another with 
metallicity increasing with decreasing age. The later two groups are  
similar to those  found for Galactic GCs \citep[and references therein]{Marin2009, Forbes2010}. 
{Studies of Galactic GCs show that the group of GCs with very old ages and 
a flat age--[Fe/H] relation are probably formed {\it in situ} in a rapid enrichment process on a 
time-scale less than 1\,Gyr. The other group of GCs  
with range of ages and an age–metallicity relation 
are probably accreted from dwarf galaxies, such as the Sgr and Canis Major (CMa)
dwarf galaxies (\citealt{Forbes2010} and references therein)}. The scenario may apply to M31. 
The difference is that those young and intermediate-age GCs in M31 are  
younger and more meta-rich than those found in the Milky Way, 
suggesting that the major accretion and merger in 
M31 continued several  (about 4; cf. Fig.\,16) Gyr even 
after that had ceased in the Milky Way.   
Most of those old, metal-rich ([Fe/H] $>~-$ 0.7\,dex)
clusters in the first group are found in the thin disk of M31 and have 
ages about 8--9\,Gyr. These cluster may also form in a rapid enrichment 
process but at a later epoch than that for the old clusters found in the halo of M31.

\subsection{Spatial distribution and radial metallicity gradient}

\begin{figure}
  \centering
  \includegraphics[width=0.48\textwidth]{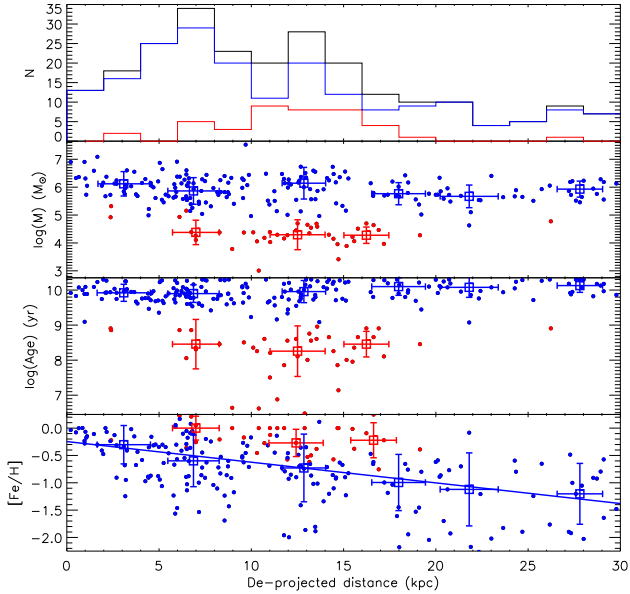}
  \caption{Distributions (top panel) 
of de-projected distances from the centre of M31 for all (black), 
young (red) and old (blue) clusters in our sample. The bottom 3 panels plot respectively the   
metallicities, ages and masses against the de-projected distances. Again,  
red and blue symbols denote young and old clusters
in our sample, respectively. Open squares with error bars give the 
average values and standard deviations in the individual 
de-projected distance bins. The blue straight line in the bottom panel 
is a linear fit of the mean metallicity values.}
  \label{famspa}
\end{figure}

The top panel of Fig.~\ref{famspa} shows the distributions 
of de-projected distances from the centre of M31 for all, young and old 
clusters in our sample.
The distribution of all clusters (dominated by old ones) peaks at $\sim$ 7\,kpc. 
The distribution of young clusters is quite different. 
They occur  from a few kpc to as far as 25\,kpc. 
The distribution has a broad weak peak around 12\,kpc. 
Some of these young clusters are clearly associated 
with the well-known star formation region, the 10\,kpc ``ring of fire'', in the M31 disk 
\citep{Brinks1984, Dame1993, Pagani1999}.

The lower three panels of Fig.~\ref{famspa} plot the masses, ages and 
metallicities of clusters in our sample against the de-projected 
distances from the centre of M31. 
Both  ages and masses of young and old clusters have a flat distribution in distance. However,
the metallicities of both old and young clusters show clear evidence 
of variations as a function of distance. The old clusters show a clear negative metallicity gradient,
measured $\Delta {\rm [Fe/H]}/\Delta R = -0.038 \pm0.023$\,dex$\rm \, kpc^{-1}$ 
by  a linear fit to the means in the individual  distance bins.
This gradient is steeper than previous findings in the literature. 
More recently, \citet{Gregersen2015} obtain 
a metallicity gradient of $-0.020 \pm 0.004 \rm \,dex\,kpc^{-1}$ 
from RGB stars  observed in the PHAT program.
In contrast to old clusters, the metallicities of young clusters
do not show clear trend of variations.
{Their mean metallicity first decreases from a value of  
$-0.2$\,dex at $7$\,kpc to $-$0.5\,dex at 13\,kpc but then  but then flattens..}

\section{Summary}

We select from Paper~I a sample of 305 massive star clusters 
in M31 and one probably associated with M33 
observed with  LAMOST since June, 2014. 
From the LAMOST  spectra combined 
with archival  $ugriz$ optical photometric measurements, we present 
new homogeneous estimates of the metallicities,
ages and masses of these clusters. 

Using the full spectral fitting code ULySS in combined 
with different SSP models (PEGASE-HR, Vazdekis et al.), 
we have determined parameters including metallicities and ages
of all clusters,  young and old. 
Values  derived  with different SSP models are in
consistent with each other for most clusters. 
Metallicities of young clusters estimated with the PEGASE-HR models 
are in better agreement with previous determinations  than those estimated with 
the models of  Vazdekis et al. In addition, ULySS fitting with the models of Vazdekis et al. 
may have incorrectly assigned a constant age, 15\,Gyr, the upper limit of the models, to
some old clusters.  Thus in general, we believe that
the ULySS code combined with the PEGASE-HR models work  better for cluster
parameter estimates than with the models of Vazdekis et al. 
For old ($t>1\,$Gyr) clusters, parameters determined with  full spectral fitting are in 
good agreement with those derived from the Lick line indices.
For young  ($t<1\,$Gyr) clusters, the spectral fitting may  have
underestimated the cluster metallicities by as much as  0.5\,dex and 
systematically overestimated the cluster ages.
We apply a SED fitting method to calculate the ages of  young clusters 
by comparing their SDSS $ugriz$ photometric data with the predictions of 
the BC03 models. The SED fitting is able to break the age-metallicity degeneracy. Ages of 
young clusters from SED fitting are in good agreement with results in the literature.
{Overall, among all the resultant parameters from those different methods, 
we prefer the metallicities from the full spectral fitting with the PEGASE-HR models for 
all objects in our sample. For the ages, we prefer those from the full spectral fitting 
with the PEGASE-HR models for old clusters and those from the multi-band SED fitting for 
young clusters.}
Based on the PEGASE-HR metallicities and ages, and 
the SED fitting ages for young clusters when available,  
we estimate the masses of 299 star clusters in our sample 
by comparing their photometry to the BC03 models. Our estimated 
metallicities, ages and masses are all in good agreement with 
previous determinations in general. 

In our sample, we find {46 young and 260 old clusters}. 
The masses determined range from $10^3$ to $10^7\,\rm M_\odot$,
peaking at $10^{4.3}\,\rm M_\odot$ and $10^{5.7}\,\rm M_\odot$ 
for young and old clusters, respectively. 
The metallicities  of M31 star clusters do
not show a clear bimodal nor a single Gaussian distribution. 
The old clusters have a  peak at [Fe/H]$=-$0.7\,dex, more metal-rich than 
the peak value of Galactic GCs. 
Young clusters have metallicities comparable to the Sun and  
clump in a small area to  the bottom-right in the age-metallicity diagram. 
The  old clusters in M31 seem to fall into three groups in the diagram.
The first group has the oldest ages and metallicities  poorer than 
$\sim -0.7$\,dex, with  no obvious age–metallicity relation. 
They were probably formed {\it in situ} in the halo in the early epoch of M\,31 
with a rapid process. The second group has metallicities also poorer than $\sim -$ 0.7\,dex but  
shows a clear age--metallicity relation, with young clusters being more metal rich than older ones. 
They probably come from disrupted dwarf galaxies 
accreted by M31 in the past. Compared to those  Milky 
Way GCs that are also believed to be accreted, 
they are younger and more metal-rich, suggesting that M31 
may have been subjected to substantial 
merger events more recently than the Milky Way. The third group of clusters has
metallicities richer than $\sim -0.7$\,dex and spans a wide range of ages. 
A significant fraction of them has ages about 8--9\,Gyr and are mainly found in the disk of 
M31. These clusters might also form {\it in situ}  in the disk of M31, 
but at an epoch much  later than those formed in the halo.

Most of the young clusters fall at de-projected distances between 
7 and 17\,kpc to the centre of M31.
Young clusters found near de-projected distance of 13\,kpc, 
probably associated  with the ring structure of M31,
have lower metallicities than the other young clusters. 
Finally, old clusters have a wide spatial distribution, peaking around 7\,kpc.
Those  in the inner region (of de-projected distances 0 -- 30\,kpc) show a
clear metallicity gradient 
of $\Delta {\rm [Fe/H]}/\Delta R = -0.038\pm0.023$\,dex$\rm \, kpc^{-1}$.

Some of the star clusters in our sample are associated with the 
streams in the outer halo of M31. Among them, LAMOST-1, 
MCGC8 and H26 are probably associated with the Giant Stellar Stream,
Stream D and Stream C, respectively. Metallicities determined for  these objects 
in the current work are consistent with this interpretation.
Most of the distant  clusters in our sample are very old except for LAMOST-1, 
and are therefore most likely relics of  early-type dwarf galaxies accreted by M31. 

\acknowledgements{
{We want to thank the referee for detailed and constructive comments that 
help improve the manuscript significantly.}
This work is partially supported by National Key Basic Research Program of China
2014CB845700 and  China Postdoctoral Science Foundation 2014M560843. 
The LAMOST FELLOWSHIP is supported by Special Funding for Advanced Users,
budgeted and administrated by Center for Astronomical Mega-Science, Chinese 
Academy of Sciences (CAMS). B.Q.C thanks Professor Jun Ma for very useful comments.
Z.F. is supported by the National Natural Science Foundation of China (NFSC) 11373003 and
National Key Basic Research Program of China (973 Program) No. 2015CB857002.
G.C.L. is supported by the National Natural Science Foundation of China (NFSC) 11303020.

This work has made use of data products from the Guoshoujing Telescope (the
Large Sky Area Multi-Object Fibre Spectroscopic Telescope, LAMOST). LAMOST
is a National Major Scientific Project built by the Chinese Academy of
Sciences. Funding for the project has been provided by the National
Development and Reform Commission. LAMOST is operated and managed by the
National Astronomical Observatories, Chinese Academy of Sciences.

Funding for SDSS-III has been provided by the Alfred P. Sloan Foundation, the 
Participating Institutions, the National Science Foundation, and the U.S. Department 
of Energy Office of Science. The SDSS-III web site is http://www.sdss3.org/.

SDSS-III is managed by the Astrophysical Research Consortium for the 
Participating Institutions of the SDSS-III Collaboration including the University 
of Arizona, the Brazilian Participation Group, Brookhaven National Laboratory, 
Carnegie Mellon University, University of Florida, the French Participation Group, 
the German Participation Group, Harvard University, the Instituto de Astrofisica de 
Canarias, the Michigan State/Notre Dame/JINA Participation Group, Johns Hopkins 
University, Lawrence Berkeley National Laboratory, Max Planck Institute for 
Astrophysics, Max Planck Institute for Extraterrestrial Physics, New Mexico 
State University, New York University, Ohio State University, Pennsylvania State 
University, University of Portsmouth, Princeton University, the Spanish Participation 
Group, University of Tokyo, University of Utah, Vanderbilt University, University of 
Virginia, University of Washington, and Yale University.}

\bibliographystyle{aa}
\bibliography{gcam}
\end{document}